\documentclass[acmsmall,nonacm]{acmart}
\usepackage{preamble}
\newcommand{\irtovec}{{\textsc{IR2Vec}}\xspace}

\makeatother

\AtBeginDocument{%
  \providecommand\BibTeX{{%
    \normalfont B\kern-0.5em{\scshape i\kern-0.25em b}\kern-0.8em\TeX}}}

\begin{document}
\title{\irtovec: LLVM IR based Scalable Program Embeddings}

\author{S. VenkataKeerthy}
\email{cs17m20p100001@iith.ac.in}
\orcid{0000-0003-1393-7321}
\author{Rohit Aggarwal}
\email{cs18mtech11030@iith.ac.in}
\author{Shalini Jain}
\email{cs15resch11010@iith.ac.in}
\author{Maunendra Sankar Desarkar}
\email{maunendra@cse.iith.ac.in}
\author{Ramakrishna Upadrasta}
\email{ramakrishna@cse.iith.ac.in}
\affiliation{%
  \institution{Indian Institute of Technology Hyderabad}
  \city{Hyderabad}
  \postcode{502 285}
}

\author{Y. N. Srikant}
\email{srikant@iisc.ac.in}
\affiliation{%
  \institution{Indian Institute of Science}
  \city{Bangalore}
  \postcode{560 012}
}

\renewcommand{\shortauthors}{VenkataKeerthy, et al.}

\begin{abstract}
We propose \irtovec, a Concise and Scalable encoding infrastructure to represent programs as a distributed embedding in continuous space. 
This distributed embedding is obtained by combining representation learning methods with flow information to capture the syntax as well as the semantics of the input programs. 
As our infrastructure is based on the Intermediate Representation (IR) of the source code, obtained embeddings are both language and machine independent.
The entities of the IR are modeled as relationships, and their representations are learned to form a \textit{seed embedding vocabulary}. 
Using this infrastructure, we propose two incremental encodings: \textit{Symbolic} and \textit{Flow-Aware}. \textit{Symbolic} encodings are obtained from the \textit{seed embedding vocabulary}, and \textit{Flow-Aware} encodings are obtained by augmenting the \textit{Symbolic} encodings with the flow information.

We show the effectiveness of our methodology on two optimization tasks (Heterogeneous device mapping and Thread coarsening). 
Our way of representing the programs enables us to use non-sequential models resulting in orders of magnitude of faster training time. 
Both the encodings generated by \irtovec outperform the existing methods in both the tasks, even while using \textit{simple} machine learning models. 
In particular, our results improve or match the state-of-the-art speedup in 11/14 benchmark-suites in the device mapping task across two platforms and 53/68 benchmarks in the Thread coarsening task across four different platforms.
When compared to the other methods, our embeddings are \textit{more scalable}, \textit{is non-data-hungry}, and \textit{has better Out-Of-Vocabulary (OOV) characteristics}.
\end{abstract}



\maketitle

\section{Introduction}

With the growth of computing, comes the growth in computations. These computations are necessarily the implementation of well-defined algorithms~\cite{Cormen:2009:IAT:1614191} as programs.
Running these programs on the rapidly evolving diverse architectures poses a challenge for compilers (and optimizations) for exploiting the best \textit{performance}.
Because most of the compiler optimizations are NP-Complete or undecidable~\cite{muchnick1997advanced,Rice-10.2307/1990888}, most of the modern compilers use carefully hand-written heuristics for extracting superior performance of these programs on various architectures.

Several attempts have been made to improve the optimization decisions by using machine learning algorithms instead of relying on sub-optimal heuristics.
Some of such works include prediction of unroll factors~\cite{Stephenson2005}, inlining decisions~\cite{Simon-Inlining-10.1109/CGO.2013.6495004}, determining thread coarsening factor~\cite{magni2014automatic}, Device mapping~\cite{o2013portable-grewe}, vectorization~\cite{Mendis2019,HajAli2020} etc.  
For such optimization applications, it is crucial to extract information from programs so that it can be used to feed the machine learning models to drive the optimization decisions. 
The extracted information should be in a form that is \textit{amenable to learning} so as to improve the optimizations on the input programs.

Primarily, there are two ways of representing programs as inputs to such machine learning algorithms -- Feature-based representations and Distributed representations.
Feature-based representation involves representing programs using hand-picked features---designed by domain experts~\cite{Fursin2011, o2013portable-grewe, magni2014automatic}---specific for the particular downstream applications. Examples for the features could be the number of basic blocks, number of branches, number of loops, and
even the derived/advanced features like arithmetic intensity. On the other hand, representation learning involves using a machine learning model to \textit{automatically learn} to represent the input as a distributed vector~\cite{Bengio:2013}. This learned representation---encoding of the program---is often called as \textit{program embedding}.
Such a distributed representation is a real-valued vector whose dimensions cannot be distinctly labelled, as opposed to that of feature-based representations. 

However, most of the existing works on using distributed learning methods to represent programs use some form of Natural Language Processing for modeling; all of them exploit the statistical properties of the code and adhere to the \textit{Naturalness hypothesis}~\cite{allamanis2018survey}. These works primarily use Word2Vec methods like skip-gram and CBOW~\cite{mikolov2013word2vec}, or encoder-decoder models like Seq2Seq~\cite{seq2seq} to encode programs as distributed vectors. 

It can be noted that most of the existing representations of programs have been designed for software engineering based applications. 
This includes algorithm classification~\cite{tbcnn-aaai16, who-wrote-this-code}, code search and recommendation~\cite{source-forager, sourcerer,Camronero2019CodeSearch, Luan2019Aroma}, 
code synthesis~\cite{abstract-syn-net}, 
Bug detection~\cite{wang2016bugram},
Code summarization~\cite{iyer2016summarizing}, 
and Software maintenance~\cite{Allamanis:learning-nat-coding-conv, Allamanis:suggesting-acc-meth-names, conv-attn-net-extreme-summ, gupta2017deepfix}.
We, however, believe that a carefully designed \textit{embedding} that can \textit{encode} the semantic characteristics of the program can be highly useful in making optimization decisions, in addition to being applied for software engineering.

In this paper, we propose \irtovec, an agglomerative approach for constructing a continuous, distributed vector to represent source code at different (and increasing) levels of IR hierarchy - Instruction, Function and Program. The vectors that are formed lower down the (program abstraction) hierarchy is used to build the vectors at higher levels. 

We make use of the LLVM compiler infrastructure~\cite{Lattner:2004:llvm} to process and analyze the code.
The input program is converted to LLVM-Intermediate Representation (LLVM-IR), a language and machine-independent format.
The initial vector representations of the (entities of the) IR, called seed embeddings, is learned by considering its statistical properties in a Representation Learning framework. 
Using these learned seed embeddings, hierarchical vectors for the new programs are formed.
To represent Instruction vectors, we propose two flavors of encodings: \textit{Symbolic} and \textit{Flow-Aware}. The \textit{Symbolic} encodings are generated directly from the learned representations.
When augmented with the flow analyses information, the Symbolic encodings become \textit{Flow-Aware}.

We show that the generic embeddings of \irtovec provide superior results when compared to the previous works like DeepTune~\cite{cummins2017end2end},
Magni et al.~\cite{magni2014automatic}, and
Grewe et al.~\cite{o2013portable-grewe} that were designed to solve specific tasks.
We also compare \irtovec with NCC by Ben-Nun et al.~\cite{ncc}; both have a similar motivation in generating generic embeddings using LLVM IR, though using different methodologies/techniques.

We demonstrate the effectiveness of the obtained encodings by answering the following \textbf{R}esearch \textbf{Q}uestions (RQ's) in the later sections:

\textbf{\textit{\phantomsection\label{RQ1}RQ1: How well do the seed embeddings capture the semantics of the entities in LLVM IR?}}

As the seed embeddings play a significant role in forming embeddings at higher levels of Program abstraction, it is of paramount importance that they capture the semantic meaning of the entities to differentiate between different programs. We show the effectiveness of the obtained seed embeddings in Sec.~\ref{subsection:seedEmbeddings-eval}.

\textbf{\textit{\phantomsection\label{RQ2}RQ2: How good are the obtained embeddings for solving diverse compiler optimization applications?}}
We show the richness of our embeddings by applying it for different tasks: (a) Heterogeneous device mapping, and (b) Prediction of thread coarsening factor in Sec.~\ref{subsection:dev-map} and~\ref{subsection:thread-coarsening} respectively.

\textbf{\textit{\phantomsection\label{RQ3}RQ3: How scalable is our proposed methodology when compared to other methods?}}
We discuss various aspects by which our encoding is more scalable than the others in Sec.~\ref{subsection:scalability}.  
We show that \irtovec has improved training time, and is non-data-hungry. Also, \irtovec does not encounter \textit{Out Of Vocabulary (OOV)} words. These are the words that have not been exposed during the training phase, and hence are not part of the seed embedding vocabulary, but are encountered during test/inference phase.

\textbf{Contributions:} The following are our contributions:
\begin{itemize}
    \item  We propose a unique way to map LLVM-IR entities to real-valued distributed embeddings, called a seed embedding vocabulary. 
    \item Using the above seed embedding vocabulary, we propose a Concise and Scalable encoding infrastructure to represent programs as vectors. 
    \item We propose two embeddings: \textit{Symbolic} and \textit{Flow-Aware} that are strongly based on classic program flow analysis theory and evaluate them on two compiler optimizations tasks: Heterogeneous device mapping and Thread coarsening.
    \item Our novel methodology of encodings is \textit{highly scalable} and performs better than the state-of-the-art techniques. We achieve an improved training time (upto $8000\times$ reduction), our method is \textit{non-data-hungry}, and 
    it \textit{does not encounter} Out-Of-Vocabulary (OOV) words. 
\end{itemize}

The paper is organized as follows: 
In Sec.~\ref{sec:relatedWorks}, we discuss various related works and categorize them.
In Sec.~\ref{sec:background}, we give the necessary background information.
In Sec.~\ref{sec:codeEmbeddings}, we explain the methodology for constructing the \textit{Symbolic} and \textit{Flow-Aware} encodings at various levels.
In Sec.~\ref{sec:experimentation}, we show our experimental setup followed by the discussion of results: first, we discuss the effectiveness of our seed embeddings, followed by our analysis and discussion on device mapping, and thread coarsening.
In Sec.~\ref{subsection:scalability}, we discuss our perspectives of \irtovec, focusing on training time, time to generate encodings and OOV issues.
Finally, in Sec.~\ref{sec:conclusions}, we conclude the paper.
\section{Related Works} 
\label{sec:relatedWorks}

Modeling code as a \textit{distributed vector} involves representing the program as a vector, whose individual dimensions cannot be distinctly labeled. Such a vector is an approximation of the original program, whose semantic meaning is ``distributed`` across multiple components. In this section, we categorize some of the existing works that model codes, based on their representations, the applications that they handle, and the embedding techniques that they use. 
Then, we discuss the details of some specific recent works in this theme.

\subsection{Representations, Applications and Embeddings}

\paragraph{Representations}
Programs are represented using standard syntactic formats like lexical tokens~\cite{Allamanis:suggesting-acc-meth-names, conv-attn-net-extreme-summ, cummins2017end2end}, Abstract Syntax Trees (ASTs)~\cite{Brauckmann2020Feb,path-based-rep:Alon:2018:GPR:3192366.3192412, Raychev:pred-prog-prop}, and standard semantic formats like Program Dependence Graphs~\cite{allamanis2018learning}, and abstracted traces~\cite{Henkel-FSE18-10.1145/3236024.3236085}. 
Then, a neural network model like RNN or its variants is trained on the representation to form distributed vectors. 

We use LLVM IR~\cite{LLVM-LangRef} as the \textit{base representation} for learning the embeddings in high dimensional space. To the best of our knowledge, we are the \textit{first ones} to model the entities of the IR---Opcodes, Operands and Types---in the form of relationships and to use a translation based model~\cite{transe-Bordes:2013:TEM:2999792.2999923} to capture such multi-relational data in higher dimensions.

\paragraph{Applications}
In the earlier works, the training to generate embeddings was either application-specific or programming language-specific: Allamanis et al.~\cite{Allamanis:suggesting-acc-meth-names} propose a token-based neural probabilistic model for suggesting meaningful method names in Java;
Cummins et al.~\cite{cummins2017end2end} propose the DeepTune framework to create a distributed vector from the tokens obtained from code to solve the optimization problems like thread coarsening and device mapping in OpenCL; 
Alon et al.~\cite{alon2019code2vec} propose code2vec, a methodology to represent codes using information from the AST paths coupled with attention networks to determine the importance of a particular path to form the code vector for predicting the method names in Java;
Mou et al.~\cite{tbcnn-aaai16} propose a tree-based CNN model to classify C++ programs;
Gupta et al.~\cite{gupta2017deepfix} propose a token-based multi-layer sequence to sequence model to fix common C program errors by students;
Other applications like learning syntactic program fixes from examples~\cite{Rolim:2017:LSP:3097368.3097417}, bug detection~\cite{Pradel:2018:DLA:3288538.3276517, DBLP:conf/iclr/WangSS18} and program repair~\cite{xiong:2017:ProgramRepair} model the code as an embedding in a high dimensional space followed by using RNN like models to synthesize fixes.
The survey by Allamanis et al.~\cite{allamanis2018survey} covers more such application-specific approaches.

On the other hand, our approach is more generic, and both application and programming language independent. 
We show the effectiveness of our embeddings on two optimization tasks (device mapping and thread coarsening) in Sec.~\ref{sec:experimentation}. 
We believe that the scope of our work can be extended beyond these applications, including program classification, code search, prediction of vectorization/unroll factors, etc.

\paragraph{Embedding techniques}
In encoding, entities are transformed into any numerical form amenable to learning, while in embedding, it is transformed to real-valued high dimensional vectors.
Several attempts~\cite{Bimodal:Allamanis:2015:BMS:3045118.3045344, ncc, prog-big-code} have been made to represent programs as distributed vectors in continuous space using word embedding techniques for diverse applications. 
Henkel et al.~\cite{Henkel-FSE18-10.1145/3236024.3236085} use the Word2Vec embedding model to generate the representation of a program from symbolic traces.
They generate and expose embeddings for the program~\cite{ncc,Henkel-FSE18-10.1145/3236024.3236085}, or the embeddings themselves become an implicit part of the training for the specific downstream task~\cite{alon2019code2vec,Pradel:2018:DLA:3288538.3276517}.

Our framework exposes a hierarchy of representations at the various levels of the program - Instruction, Function and Program level. Our approach is the first one to propose using seed embeddings. More significantly, we are the \textit{first ones} to use program analysis (flow-analysis) driven approaches---\textit{not} machine learning based approaches---to form the agglomerative vectors of programs beginning from the base seed encodings.

\subsection{Similar works}

The closest to our work is Ben-Nun et al.'s Neural Code Comprehensions (NCC)~\cite{ncc}, who represent programs using LLVM IR. They use skip-gram model~\cite{skipgram-NIPS2013_5021} on conteXtual Flow Graph (XFG), which models the data/control flow of the program to represent IR. The skip-gram model is trained to generate embeddings for every IR instruction (inst2vec).  So as to avoid Out Of Vocabulary (\textit{OOV}) statements, they maintain a large vocabulary, one which uses large ($>640M$) number of XFG statement pairs. A more thorough comparison of our work with NCC~\cite{ncc} (along with DeepTune~\cite{cummins2017end2end}) is given in Sec.~\ref{subsection:scalability}.

The recent work by Brauckmann et al.~\cite{Brauckmann2020Feb} represents programs as Abstract Syntax Trees and annotates it with the control and data flow edges. A Graph Neural Network is used to learn the representation of the program in a supervised manner to solve the specific tasks of device mapping and thread coarsening. We achieve better performance on both the tasks, whereas they fail to show improvements in the prediction of the thread coarsening factor.

Another recent work is ProGraML~\cite{Cummins2020Mar}, which constructs a flow graph with data and control flow information from the Intermediate Representation of the program. They use inst2vec embeddings~\cite{ncc} to represent the nodes of the graph, and use a Gated Graph Neural Network to represent the program. 
\section{Background} 
\label{sec:background}

\subsection{LLVM and Program semantics}
LLVM is a compiler infrastructure that translates source-code to machine code by performing various optimizations on its Intermediate Representation (LLVM IR)~\cite{Lattner:2004:llvm}. LLVM IR is a typed, well-formed, low-level, \textit{Universal IR} to represent any high-level language and translate it to a wide spectrum of targets~\cite{LLVM-LangRef}. Being a successful compiler, LLVM provides easy access to existing control and data flow analysis and lets new passes (whether analyses or transformations) to be added seamlessly. 

The building blocks of LLVM IR include Instruction, Basic block, Function and Module.
Every instruction contains opcode, type and operands, and each instruction is statically typed. A basic block is a maximal sequence of LLVM instructions without any jumps. A collection of basic blocks form a function, and a module is a collection of functions. This hierarchical nature of LLVM IR representation helps in obtaining embeddings at the corresponding levels of the program. 
Characterizing the flow of information which flows into (and out of) each basic block constitutes the data flow analysis. We study the impact of using such flow information as a part of the encodings.

\subsection{Representation Learning}
\label{subsec:rep-learning}
The effectiveness of a machine learning algorithm depends on the choice of data representation and on the specific features used. As discussed earlier, Representation Learning is a branch of machine learning that learns the representations of data by automatically extracting the useful features~\cite{Bengio:2013}. 
Unsupervised models for learning distributed representations broadly fall under two major categories:
\begin{enumerate}
    \item Context-window based embedding: Methods such as Word2Vec~\cite{mikolov2013word2vec}, GloVe~\cite{pennington2014glove} fall under this category. 
    \item Knowledge graph embedding: Methods such as TransE~\cite{transe-Bordes:2013:TEM:2999792.2999923}, TransR~\cite{Lin:2015}, TransD \cite{ji-etal-2015-knowledge}, TransH~\cite{Wang:2014} fall under this category.
\end{enumerate}

The context-window-based models operate on the basis of the surrounding context. The Knowledge graph embedding methods guide the learning of the entity representations based on the relationships that they participate in.

For program representations, we feel that it is important to consider semantic relationships rather than considering surrounding contexts; the latter could just mean the neighboring instructions. 
So, we prefer knowledge graph embedding models over context window embedding approaches. These knowledge graph embedding models use relationships to group similar datapoints together.
\vspace*{-0.1cm}
\subsubsection*{Knowledge Graph Representations} 
A Knowledge Graph (KG) is a collection of (a) entities, and (b) relationships between pairs of entities.  
Let $\langle h,r,t \rangle$ be a triplet from the knowledge graph, where the entities $h$ and $t$ are connected by relationship $r$, and  their representations are learned as \textit{translations} from head entity $h$ to the tail entity $t$ using the relation $r$ in a high-dimensional embedding space.
For the same reason, these models are termed \textit{translational}.

The representations are nothing but the vectors of pre-defined dimensions that are determined automatically by a learning method. The learning method is based on the principle that: given (representations of) any two items from the triplet (like $\langle h,r \rangle$, $\langle r,t\rangle$, $\langle t,h\rangle$), it should be possible to compute/predict the third item.  Based on the above principle, different representation learning algorithms for knowledge graphs model the relationships between $h$, $r$, and $t$ in the triplets in different ways. 

\subsubsection*{Using TransE}
Of the many varieties available for KG embeddings, we use TransE~\cite{transe-Bordes:2013:TEM:2999792.2999923}, a \textit{translational representation learning} model, which embeds $h$, $r$ and $t$ on to the same high dimensional space. It tries to learn the representations using the relationships of the form $h+r \approx t$, for a triplet $\left < h, r, t \right >$. 

This is achieved by following a \textit{margin based ranking} loss $\mathcal{L}$, where the distance between $(h+r, t)$ and $(h'+r, t')$ (where, $\langle h', r, t' \rangle$ are invalid triplets) 
is at least separated by a margin $m$. 
It is given as:

\begin{equation*}
    \mathcal{L} = \sum_{\langle h,r,t \rangle}\sum_{\langle h',r,t' \rangle} \left[m + distance(h+r,t) - distance(h'+r,t') \right]_+
\end{equation*}

Here, $[x]_+$ denotes the hinge loss.

Among the other Knowledge Graph models, TransE has relatively much smaller number of parameters to be learned, and has been used successfully in various algorithms for learning representations.  Adopting this for our setting ensures that the representations can be learned in a faster and effective manner that scales well on huge datasets~\cite{ji-etal-2015-knowledge, han2018openke}.

\subsection{Necessity for a LLVM based embedding for optimizations}
Traditionally, machine independent compiler optimizations were always considered to be part of the ``middle-end''\cite{muchnick1997advanced} of compilers. The most successful design of this language and architecture-independent representations has been the LLVM IR~\cite{Lattner:2004:llvm, LLVM-LangRef}. Using the LLVM-infrastructure, carefully well-crafted heuristics have been implemented in various optimization passes to help in achieving better performance.
As the power of using ML in compilers has been recognized~\cite{10.1023/A:1015729001611}, many ML-driven approaches have also been  proposed~\cite{DBLP:books/crc/CRCcompiler2007/Vaswani07} as alternatives to these heuristics. However, they have primarily been feature-based, which means that integrating them into the middle-end is prone to challenges~\cite{Fursin2011}, both at modeling as well as engineering level. 

Hence, there is a necessity for a \textit{language and architecture agnostic embedding based compiler framework} that bridges the gap between the need for ML-based compiler optimizations, and their adoptability in compiler infrastructures. Our work tries to bridge this particular gap between embeddings from representation learning and the LLVM compiler. We believe that our work has a unique potential to be called as a \textit{generic compiler-based embedding}. 
\section{Code Embeddings} 
\label{sec:codeEmbeddings}

In this section, we explain our methodology for obtaining code embeddings at various hierarchy levels of the IR. 
We first give an overview of the methodology and then describe the process of embedding instructions and basic blocks (BB) by considering the program flow information to form a cumulative \textit{BB vector}.
We then explain the process to represent the functions and modules by combining the individual BB vectors to form the final \textit{Code Vector}.
We propose two different incremental embeddings at the instruction level.

\subsection{Overview}
\begin{figure}[t]
    \centering
    \includegraphics[scale=0.5]{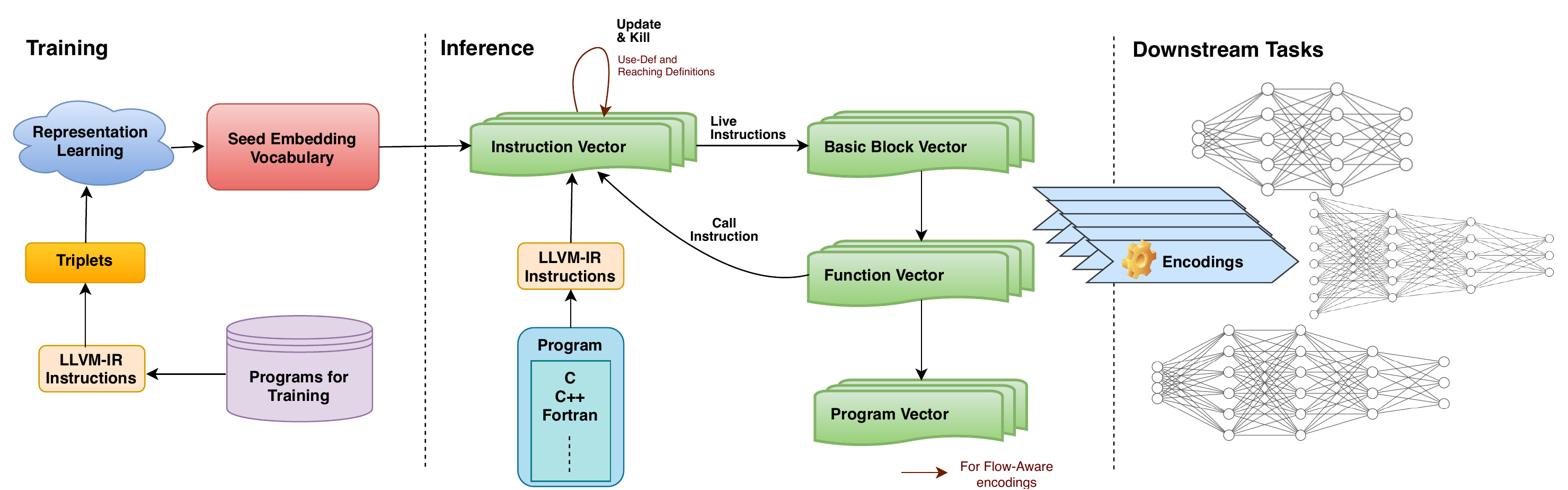}
    \caption{Overview of \irtovec infrastructure}
     \label{fig:IR2Vec-Overview}
\end{figure}

The overview of the proposed methodology is shown in Fig.~\ref{fig:IR2Vec-Overview}. Instructions in IR can be represented as an Entity-Relationship Graph, with the instruction entities as nodes, and the relation between the entities as edges. A translational learning model that we discussed in Sec.~\ref{subsec:rep-learning} is used to learn these relations (Sec.~\ref{subsec:preprocessing}). The output of this learning is a dictionary containing the embeddings of the entities and is called \textit{Seed embedding vocabulary}. 

The above dictionary is looked up to form the embeddings at various levels of the input program. 
At the coarsest level, instruction embeddings are obtained just by using the \textit{Seed embedding vocabulary}. We call such encodings as \textit{Symbolic} encodings.
We use the \textit{Use-Def} and \textit{Reaching definition}~\cite{Hecht:1977:FAC:540175, muchnick1997advanced} information to form the instruction vector for \textit{Flow-Aware} encodings. 

The instructions which are \textit{live} are used to form the \textit{Basic block Vector}. This process of formation of a basic block vector using the flow analysis information is explained in Sec.~\ref{subsec:DFinfo-prop}. The vector to represent a function is obtained by using the basic block vectors of the function. The \textit{Code vector} is obtained by propagating the vectors obtained at the function level with the \texttt{call graph} information, as explained in Sec.~\ref{subsec:CFInfo-prop}.

\subsection{Learning Seed Embeddings of LLVM IR}
\label{subsec:preprocessing}
\subsubsection{Generic tuples}
The opcode, type of operation (int, float, etc.) and arguments are extracted from the LLVM IR. This extracted IR is preprocessed in the following way: 
first, the identifier information is abstracted out with more generic information, as shown in Tab.~\ref{tab:instruction-mapping}.
Next, the Type information is abstracted to represent a base type ignoring its width. For example, the type \texttt{i32} of LLVM IR is represented as \texttt{int}.
\vspace*{-0.4cm}
\begin{table}[h]
  \caption{Mapping identifiers to generic representation}
  \vspace*{-\baselineskip}
  \label{tab:instruction-mapping}    \small
  \begin{tabular}{ll}
    \toprule
     \textbf{Identifier} & \textbf{Generic representation} 
     \\
        \hline
        Variables &  \texttt{VAR} \\
        Pointers &  \texttt{PTR} \\
        Constants &  \texttt{CONST} \\
        Function names & \texttt{FUNCTION}\\
        Address of a basic block &  \texttt{LABEL} \\
\bottomrule
\end{tabular}
\vspace*{-0.4cm}
\end{table}

\subsubsection{Code triplets}
\begin{figure}[t]
    \hspace*{-0.7cm} 
    \includegraphics[scale=0.9]{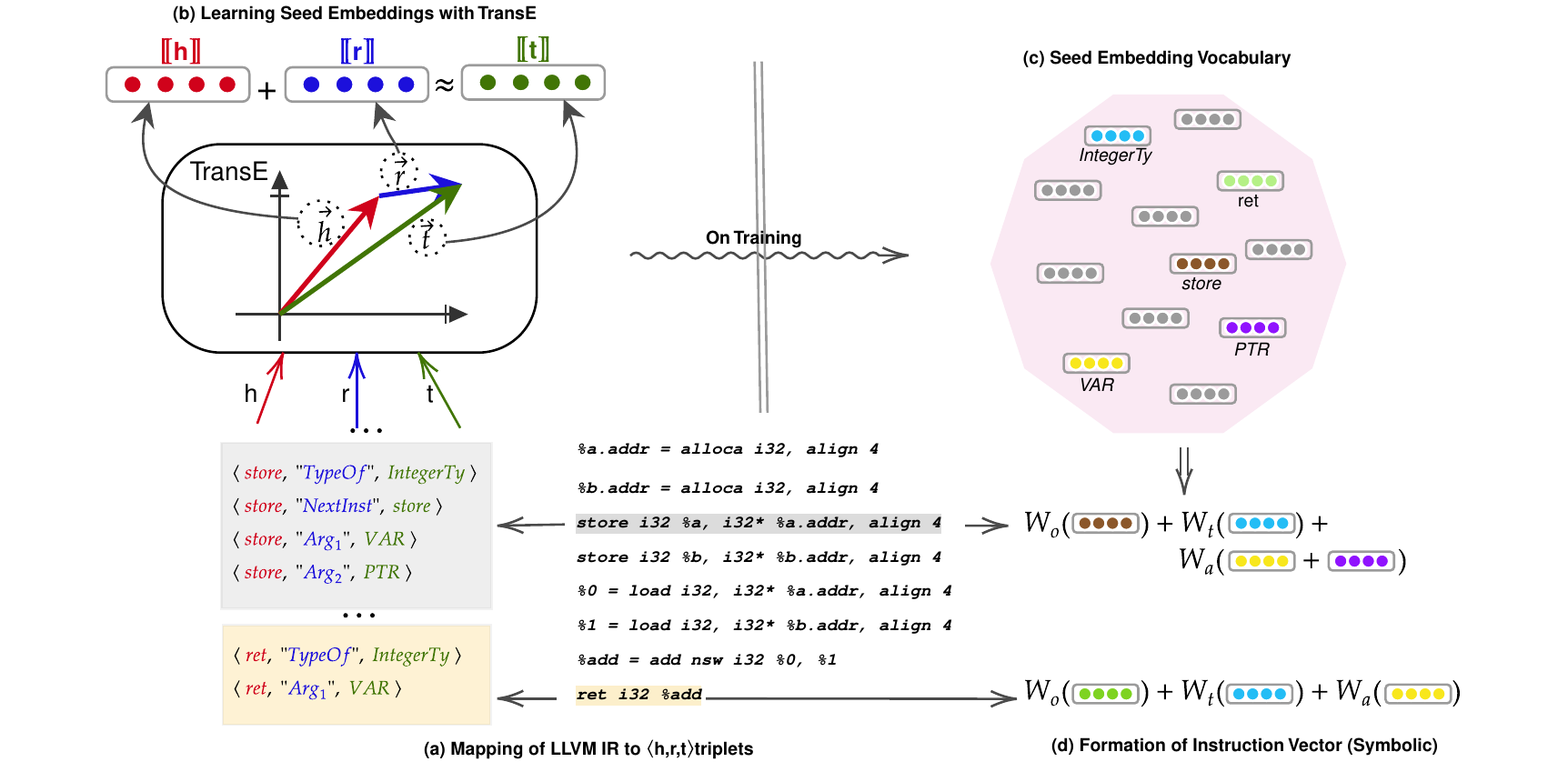}
    \caption{Overall schematic of \irtovec: 
    (a) [Data Collection] Mapping LLVM IR to $\langle h, r, t \rangle$ triplets.  
    (b) [Vocabulary training] Learning representations using TransE.  
    (c) [After Training] Obtained Seed Embedding Vocabulary.
    (d) [Inference] Formation of instruction vectors for new programs.
    }
    \label{fig:llvm-ir-example}
     \vspace*{-0.5cm}
\end{figure}

From this preprocessed data, three major relations are formed: (1) \texttt{TypeOf}: Relation between the opcode and the type of the instruction, (2) \texttt{NextInst}: Relation between the opcode of the current instruction and opcode of the next instruction; (3) \texttt{Arg$_i$}: Relation between opcode and its i{$^{th}$} operand. This transformation from actual IR to relation ($\langle h, r, t \rangle$ triplets) is shown in Fig.~\ref{fig:llvm-ir-example}(a). These triplets form the input to the representation learning model. 

\paragraph{Example}
The LLVM IR shown in Fig.~\ref{fig:llvm-ir-example} corresponds to a function that sums up the two integer arguments and returns the result. The first \texttt{store} instruction of LLVM IR in Fig.~\ref{fig:llvm-ir-example}(a) is of integer type with a variable as the first operand and a pointer as the second operand. It is transformed into the corresponding triplets involving the relations \texttt{TypeOf, NextInst, Arg$_1$, Arg$_2$}, as shown. $\square$

\subsubsection{Learning Seed Embedding Vocabulary}
As shown in Fig.~\ref{fig:llvm-ir-example}(b), the generated Code triplets $\langle h, r, t \rangle$ are used as input to the TransE learning model (Sec.~\ref{subsec:rep-learning}).
On training, the model learns the representations of the entities forming the \textit{Seed embedding vocabulary}, as shown in Fig.~\ref{fig:llvm-ir-example}(c). 

\subsection{Instruction Vector}
\label{subsec:DFinfo-prop}
 Let the entities of instruction $l$, be represented as $O^{(l)}$, $T^{(l)}$, ${A_i}^{(l)}$ - corresponding to Opcode, Type and i$^{th}$ Argument of the instruction and their corresponding vector representations from the learned \textit{seed embedding vocabulary} be $\llbracket \mathbf{O}^{(l)} \rrbracket$, $\llbracket \mathbf{T}^{(l)} \rrbracket$, $\llbracket \mathbf{{A_i}}^{(l)} \rrbracket$.
Then, an instruction of format 
\[
\left<
O^{(l)} \, \, T^{(l)} {A_1}^{(l)}  {A_2}^{(l)} \cdots {A_n}^{(l)}
\right>
\] 
is represented as a vector which is computed as:
\begin{equation}
    \label{eqn:inst-vec}
    W_o.\llbracket \mathbf{O}^{(l)} \rrbracket + W_t.\llbracket \mathbf{T}^{(l)} \rrbracket + W_a.\left(\llbracket \mathbf{A_1}^{(l)} \rrbracket + \llbracket \mathbf{A_2}^{(l)} \rrbracket + \cdots + 
    \llbracket \mathbf{A_n}^{(l)} \rrbracket \right)
\end{equation}

Here, $W_o$, $W_t$ and $W_a$ are scalars ($\in [0, 1]$), the plus ($+$) denotes the element-wise vector addition operator, and the dot ($.$) denotes the scalar multiplication operator. Further, the $W_o$, $W_t$ and $W_a$ are chosen with a heuristic that gives more weightage to opcode than type, and more weightage to type than arguments: 
\begin{equation}
\label{eqn:heuristic}
W_o > W_t > W_a
\end{equation}

This resultant vector that represents an instruction is the \textit{Instruction vector} in \textbf{Symbolic encodings}. 

\paragraph{Example (contd.)}
For the \texttt{store} instruction shown in Fig.~\ref{fig:llvm-ir-example}(d), the representations of opcode \texttt{store}, type \textit{IntegerTy}, and arguments \textit{VAR}, \textit{PTR} are fetched from the seed embedding vocabulary, and the instruction is represented as $W_o.(\llbracket \mathbf{store} \rrbracket) + W_t.(\llbracket \mathbf{IntegerTy} \rrbracket) + W_a.(\llbracket \mathbf{VAR} \rrbracket + \llbracket \mathbf{PTR} \rrbracket)$. 
In the same figure, we also show a similar example of the \texttt{return} instruction.  $\square$

\subsubsection{Embedding Data flow information}
An instruction in LLVM IR may define a variable or a pointer that could be used in another section of the program. The set of uses of a variable (or pointer) gives rise to the \textit{use-def} (UD) information of that particular variable (or pointer) in LLVM IR~\cite{Hecht:1977:FAC:540175, muchnick1997advanced}.
In imperative languages, a variable can be redefined; meaning, in the flow of the program, it has a different set of lifetimes. Such a redefinition is said to \textit{kill} its earlier definition. During the flow of program execution, only a subset of \textit{live} definitions would reach the use of the variable.
Those definitions that \textit{reach} the use of a variable are called its \textit{Reaching Definitions}.

We model the instruction vector using such flow analyses information to form \textit{\textbf{Flow-Aware}} encodings. 
Each $A_j$, which has been already defined, is represented using the embedding of its \textit{reaching definitions}. The Instruction Vector for a reaching definition, if not calculated, is computed in a demand-driven manner.

\subsubsection{Instruction Vector for Flow-Aware encodings}
\label{subsec:CFInfo-prop}
If $RD_1$, $RD_2$, ... , $RD_n$ are the reaching definitions of ${A_j}^{(l)}$, and $\llbracket \mathbf{RD_i} \rrbracket$ be their corresponding representations, then, the encoding $\llbracket \mathbf{{A_j}}^{(l)} \rrbracket$ is calculated by aggregating over all the vectors of the reaching definitions as follows:
\begin{equation}
    \label{eqn:RD-args}
    \llbracket \mathbf{{A_j}}^{(l)} \rrbracket = \sum_{i\ =\ 1}^{n} \llbracket \mathbf{RD_i} \rrbracket
\end{equation}

The $\Sigma$ stands for the vector sum of the operands.

For the cases where the definition is not available (for example, function parameters), the generic entity representation of "VAR" or "PTR" from the learned \textit{seed embedding vocabulary} is used.

\begin{figure}[ht]
  \begin{tabular}{ll}
    \begin{minipage}{0.7\textwidth}
        \captionsetup[subfigure]{font=scriptsize,labelfont=scriptsize}
        \renewcommand{\thesubfigure}{b}
        \subfloat[\scriptsize{Example CFG}]{\includegraphics[width=0.95\textwidth]{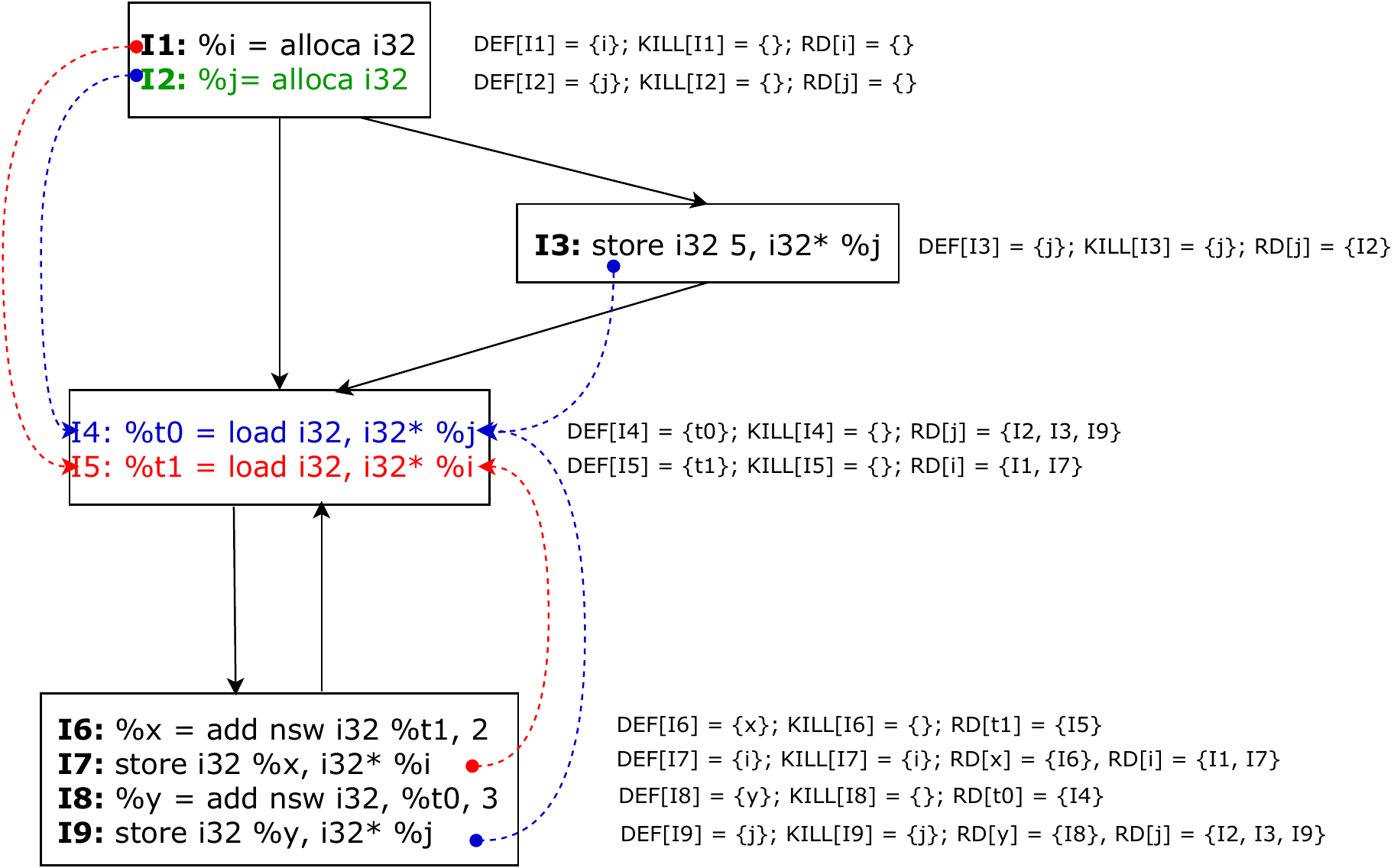}\label{fig:interbb-eg}}   
    \end{minipage}\hspace{-1.5cm} & 
    \begin{minipage}{0.7\textwidth}
        \captionsetup[subfigure]{font=scriptsize,labelfont=scriptsize}
        \hspace{0.8cm}
        \vspace{3.2cm}
        \renewcommand{\thesubfigure}{a}
        \subfloat[\scriptsize{Overview}]{\label{fig:prop-usedef} \includegraphics[width=0.4\textwidth]{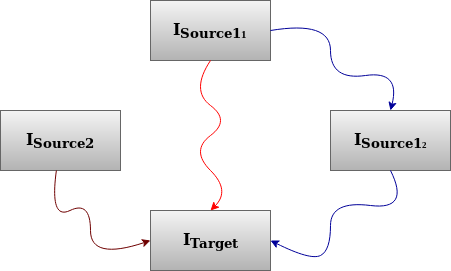}}%
        \qquad
        \hspace{-4.8cm}
    \end{minipage}\hspace{-1.5cm}
    \begin{minipage}{0.31\textwidth}
        \captionsetup[subfigure]{font=scriptsize,labelfont=scriptsize}
        \hspace{-8cm}
        \vspace{-3.2cm}
        \renewcommand{\thesubfigure}{c}
        \subfloat[\scriptsize{Instruction vectors corresponding to Fig.~\ref{fig:interbb-eg}}]{
            \scriptsize
            \fbox{%
                \parbox{\textwidth}{%
                    \centering
                    \color{olivegreen}
                    $\llbracket \mathbf{I_2} \rrbracket = W_o(\llbracket \mathbf{alloca} \rrbracket) + W_t(\llbracket \mathbf{IntegerTy} \rrbracket) $ \\
                    \vspace{0.25cm}
                    \color{blue}
                    $\llbracket \mathbf{I_4} \rrbracket = W_o(\llbracket \mathbf{load} \rrbracket) + W_t(\llbracket \mathbf{PointerTy} \rrbracket) +$\\
                    \vspace{0.18cm}\hspace*{\fill}$W_a(\llbracket \mathbf{I_2} \rrbracket + \llbracket \mathbf{I_3} \rrbracket + \llbracket \mathbf{I_9} \rrbracket)$\\
                    \vspace{0.25cm}
                    \color{red}
                    $ \llbracket \mathbf{I_5} \rrbracket = W_o(\llbracket \mathbf{load} \rrbracket) + W_t(\llbracket \mathbf{PointerTy} \rrbracket) + $\\
                    \vspace{0.18cm}\hspace*{\fill}$W_a(\llbracket \mathbf{I_1} \rrbracket + \llbracket \mathbf{I_7} \rrbracket)$\\
                }
            }
        }
    \end{minipage} \\
  \end{tabular}
\caption{Illustration of generating inter basic block Instruction Vectors.}
\label{fig:inst-vec}
\end{figure}

An illustration is shown in Fig.~\ref{fig:prop-usedef}, where the instructions $I_{Source1_1}$ and $I_{Source2}$ reach $I_{Target}$ as arguments. Here, the definition of $I_{Source1_1}$ could reach the argument $A_{Target}$ of the instruction $I_{Target}$ either directly or via $I_{Source1_2}$. And, definition $I_{Source2}$ reaches the argument $A_{Target}$ of the instruction $I_{Target}$ directly. Hence, $\llbracket \mathbf{A_{Target}} \rrbracket$ is computed as $\llbracket \mathbf{I_{Source1_1}} \rrbracket$ +  $\llbracket \mathbf{I_{Source2}} \rrbracket$.

An instruction is said to be \textit{killed} when the return value of that instruction is redefined. As LLVM IR is in SSA form~\cite{ssa:cytron1991efficiently, Lattner:2004:llvm}, each variable has a single definition, and the memory gets (re-)defined. 
Based on this, we categorize the instructions into two classes: ones which define memory, and the ones that do not. 
The first class of instructions is \textit{Write} instructions, and the second class of instructions is \textit{Read} instructions.

\begin{figure}[ht]
    \centering
    \includegraphics[scale=0.155]{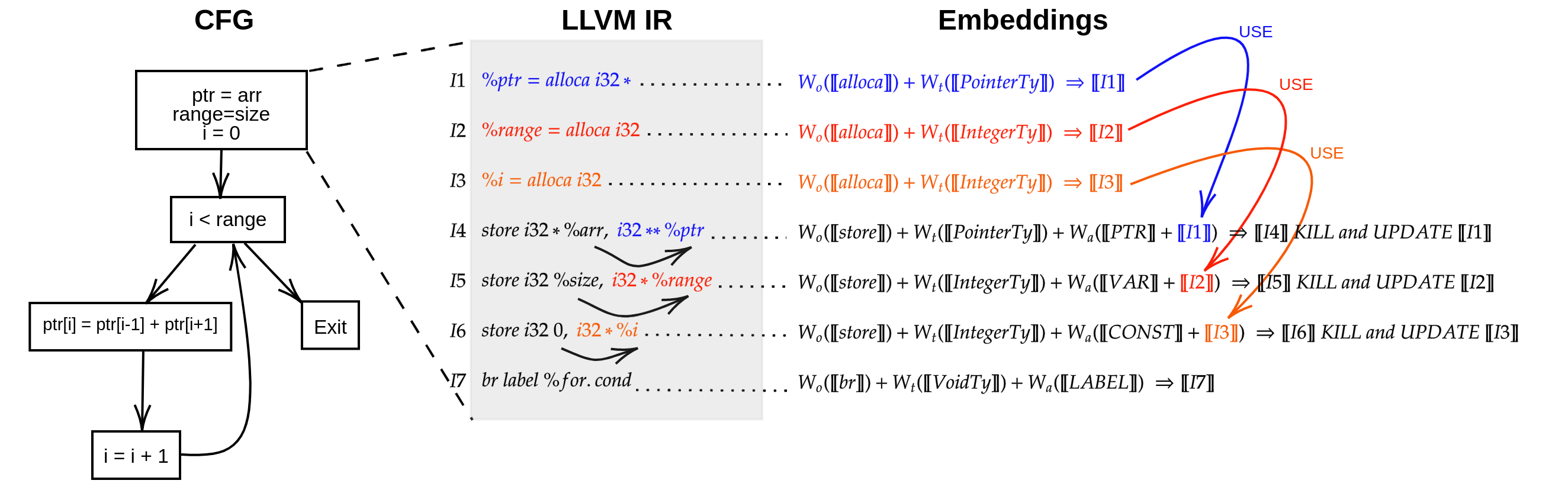}
    \caption{{Illustration of generating intra basic block Instruction Vectors}}
    \label{fig:IntraBB-dataflow}
\end{figure}

For each instruction, the embeddings are formed as explained above.  If these embeddings correspond to a write instruction, future uses of the redefined value will take the embedding of the current instruction, instead of the embedding corresponding to its earlier (killed) definition, until the current definition itself gets redefined.
This process of \textit{Kill and Update}, along with the \textit{use} of \textit{reaching definitions} for forming the instruction vectors within (and across) the basic block is illustrated in Fig.~\ref{fig:IntraBB-dataflow} (and Fig.~\ref{fig:inst-vec}) for the corresponding Control Flow Graph (CFG) respectively. 

\paragraph{Example (contd.)}
The CFG in Fig.~\ref{fig:IntraBB-dataflow} corresponds to a function that takes \texttt{arr} and \texttt{size} as its two arguments.
We expand the first basic block of this CFG and show its corresponding LLVM IR.
The values of the two arguments are allocated memory in $I1$ and $I2$; this is followed by storing them to the local variables, \texttt{ptr} and \texttt{range} by the \texttt{store} instructions in $I4$ and $I5$.
Similarly, memory for the loop induction variable \texttt{i} is allotted by $I3$ and is initialized to zero by $I6$. $\square$ 

Here, we show the process of propagating instruction vectors within the basic block.

\paragraph{Example (contd.)}
The definition of \texttt{ptr} in $I1$ reaches the use of \texttt{ptr} in $I4$. So, the representation of $I1$ is used in its place as shown, instead of the generic representation of PTR. Also, the store instruction kills the definition of \texttt{ptr} in $I1$, as it updates the value of \texttt{ptr} with that of \texttt{arr}. Hence, in the further uses of \texttt{ptr}, the representation of $I4$ is used instead of $I1$. The same is the case for other store instructions in $I5$ and $I6$. $\square$

Similarly, in Fig.~\ref{fig:inst-vec}, we show how the propagation of instruction vectors happens across basic blocks. In this case, more than one definition can reach a use via multiple control paths. Hence, as explained in Eqn.~\ref{eqn:RD-args}, all possible definitions that can potentially reach a use are considered to represent the argument of that instruction. 

\paragraph{Example (contd.)}
In Fig.~\ref{fig:inst-vec}(b), the definitions of \texttt{j} from $I2$, $I3$ and $I9$ can reach argument \texttt{j} of instruction $I4$ without being killed. Hence we conservatively consider all three definitions for representing the argument \texttt{j} in $I4$ as shown in Fig.~\ref{fig:inst-vec}(c). Fig~\ref{fig:inst-vec}(b) also shows another such example for the instruction $I5$. $\square$

This resulting instruction vector which is formed by adding the exact flow information on the \textit{Symbolic} encodings' instruction vector is used in obtaining \textit{\textbf{Flow-Aware}} encodings. 

\subsubsection{Resolving Circular Dependencies}
\label{subsubsec:ResolvingCirDep}
While forming the instruction vectors, circular dependencies between two write instructions may arise if both of them write to the same location and the (re-)definitions are reachable from each other. 

\begin{figure}[h]
    \begin{tabular}{ll}
    \begin{minipage}{0.7\textwidth}
        \includegraphics[scale=0.5]{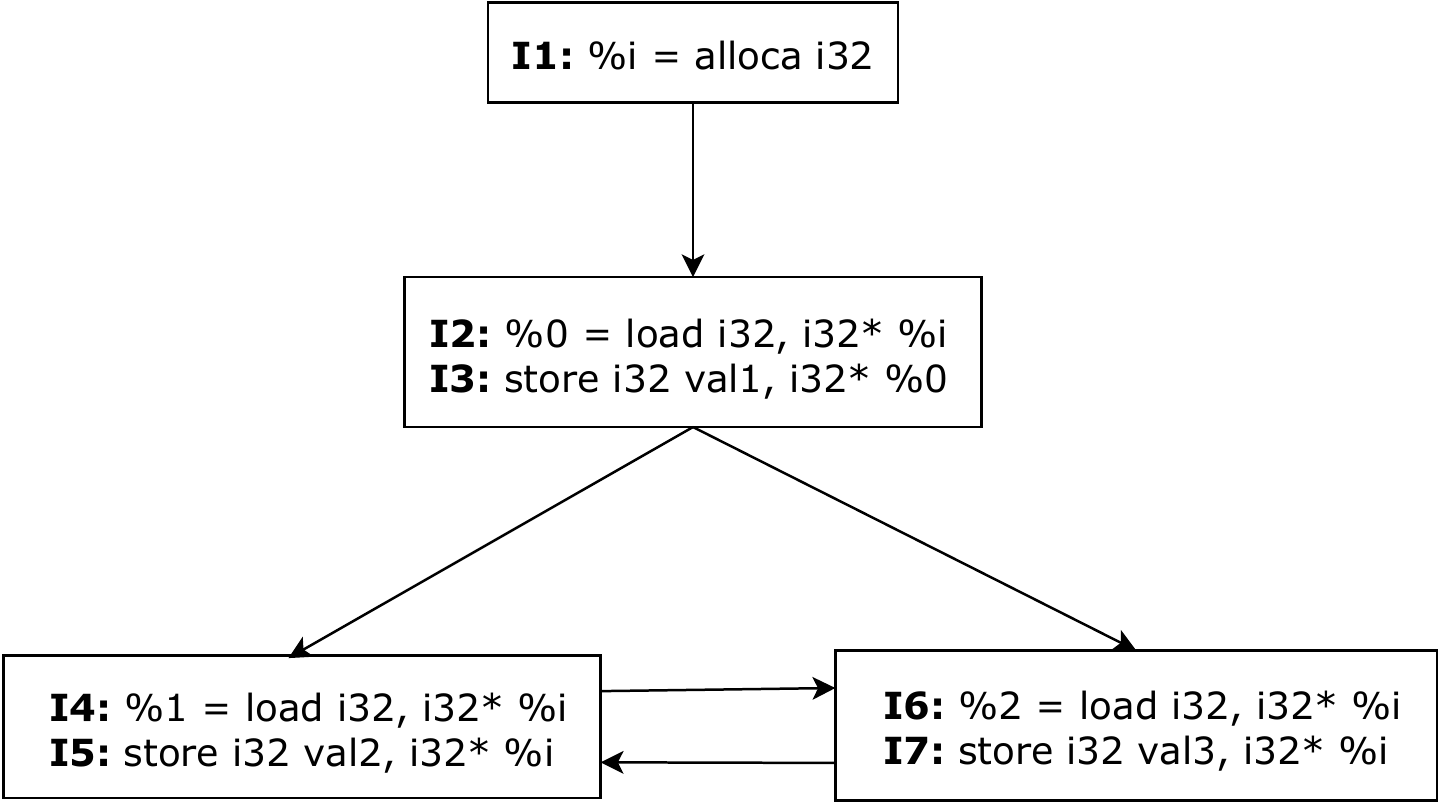}
    \end{minipage}\hspace{-4.5cm} & 
\begin{minipage}{0,6\textwidth}
\vspace{-0.5cm}
\scriptsize
\centering

\begin{displaymath}
    RD[\mathbf{I_7}]= \{\mathbf{I_3}, \mathbf{I_5}\}; \ \ 
 But \ \llbracket \mathbf{I_5} \rrbracket \ is \ not \ calculated, yet \implies Calculate \  \llbracket \mathbf{I_5} \rrbracket
\end{displaymath}
\vspace{0.25cm}
\begin{displaymath}
\begin{split}
    RD[\mathbf{I_5}]= \{\mathbf{I_3}, \mathbf{I_7}\}; \ \ 
 But \ \llbracket \mathbf{I_7} \rrbracket \ is \ not \ calculated, yet \implies Calculate \  \llbracket \mathbf{I_7} \rrbracket
 \\
 \hspace*{\fill}
 \implies \llbracket \mathbf{I_5} \rrbracket \ depends \ on \ \llbracket \mathbf{I_7} \rrbracket \ and \ vice \ versa.
\end{split}
\end{displaymath}
\end{minipage} \\
\end{tabular}
    \caption{Control Flow Graph showing circular dependency}
    \label{fig:ciculardep-cfg2}
\end{figure}

\paragraph{Example}
To calculate $\llbracket \mathbf{I_4} \rrbracket$ (the encoding of $I_4$), in the CFG shown in Fig.~\ref{fig:ciculardep-cfg2}\footnote{Though this CFG is the classic irreducibility pattern~\cite{Hecht:1977:FAC:540175, muchnick1997advanced}, it is easy to construct reducible CFGs  which have circular dependencies.}, it can be seen that the definitions of \texttt{i} from $I3$ and $I7$ can reach the argument \texttt{i} in $I4$. 
However, $\llbracket \mathbf{I_5} \rrbracket$ is needed for computing $\llbracket \mathbf{I_7} \rrbracket$ and is yet to be computed. But for computing $\llbracket \mathbf{I_5} \rrbracket$, $\llbracket \mathbf{I_7} \rrbracket$ is needed. Hence this scenario results in a circular dependency. $\square$

This problem can be solved by posing the corresponding embedding equations as a set of simultaneous equations to a solver. For example, the embedding equations of $I_5$ and $I_7$ shown in Fig.~\ref{fig:ciculardep-cfg2} would be:

\begin{equation}
    \label{eqn:ex-I5-I7}
    \begin{aligned} 
            \llbracket \mathbf{I_4} \rrbracket =& W_o(\llbracket \mathbf{load} \rrbracket) + W_t(\llbracket \mathbf{IntegerTy} \rrbracket) + W_a(\llbracket \mathbf{I_3} \rrbracket + \llbracket \mathbf{I_7} \rrbracket)
        \\
            \llbracket \mathbf{I_5} \rrbracket =& W_o(\llbracket \mathbf{store} \rrbracket) + W_t(\llbracket \mathbf{IntegerTy} \rrbracket) + W_a( \llbracket \mathbf{VAR} \rrbracket) + W_a(\llbracket \mathbf{I_3} \rrbracket + \llbracket \mathbf{I_7} \rrbracket)
        \\
            \llbracket \mathbf{I_7} \rrbracket =& W_o(\llbracket \mathbf{store} \rrbracket) + W_t(\llbracket \mathbf{IntegerTy} \rrbracket) + W_a(\llbracket \mathbf{VAR} \rrbracket) + W_a(\llbracket \mathbf{I_3} \rrbracket + \llbracket \mathbf{I_5} \rrbracket)
    \end{aligned}
\end{equation}

It can be seen that Eqn.~\ref{eqn:ex-I5-I7} is a system of linear equations, where $\llbracket \mathbf{I_4} \rrbracket$, $\llbracket \mathbf{I_5} \rrbracket$ and $\llbracket \mathbf{I_7} \rrbracket$ are the unknowns that are to be calculated, while the rest of the values are known. Such embedding equations form a system of linear equations that can be posed to a solver to find the solution.

Just like any system of linear equations, there are three cases of solutions.
\begin{enumerate}
    \item \textbf{Unique solution:} In this case, the solution is obtained in a straight forward manner.
    \item \textbf{Infinitely many solutions:} In this case, any one of the obtained solutions can be considered as the result. 
    
    \item \textbf{No solution:} In this case, to obtain a solution, we perturb the value of $W_a$ as $W_a = W_a - \delta$ so that the modified system converges to a solution, with $\delta$ chosen randomly.  
    If the system does not converge with the chosen value of $\delta$, another $\delta$ could be picked, and the process can be iterated until the system converges.
\end{enumerate}

In our entire experimentation setup described in Sec.~\ref{sec:experimentation}, we however did not encounter cases (2) and (3).

\subsection{Construction of Code vectors from Instruction vector}
After computing the instruction vector for every instruction of the basic block, we compute the cumulative Basic Block vector by using the embeddings of those instructions that are not killed. 

For a basic block $BB_i$, the representation is computed as the sum of the representations of \textit{live} instructions $LI_1, LI_2, ..., LI_m$ in $BB_i$.
\vspace{-0.33cm}
\begin{equation}
    \textbf{$\llbracket \mathbf{BB_i} \rrbracket = \sum_{k=1}^{m} \llbracket \mathbf{LI_k} \rrbracket$}
\end{equation}

\begin{procedure}[H]
\label{algorithm:inst-vec}
\DontPrintSemicolon
\SetAlgoNoLine
\footnotesize
\If{$\llbracket \mathbf{I} \rrbracket$ is already computed}{
\Return{$\llbracket \mathbf{I} \rrbracket$} 
}

$\llbracket \mathbf{O} \rrbracket\ \leftarrow$ seedEmbeddings[Opcode(I)]\tcp*{Fetch value of Opcode of I from seed embeddings}
$\llbracket \mathbf{T} \rrbracket\ \leftarrow$ seedEmbeddings[Type(I)]\tcp*{Fetch value of Type of I from seed embeddings}
$\llbracket \mathbf{A} \rrbracket\ \leftarrow \emptyset$ \tcp*{Initializing n-d Argument vector to zeroes}
\For{each argument  $A_i \in$ I}{
    \If{$A_i\ $ is a Function}{
        $\llbracket \mathbf{A_i} \rrbracket  \leftarrow$  seedEmbeddings[``FUNCTION'']\;
        \If{Encoding is \textit{Flow-Aware} and definition of $A_i$ is available}{
            $\llbracket \mathbf{A_i} \rrbracket  \leftarrow$  getFuncVec($A_i$)\;
        }
    }
    \ElseIf{$A_i$ $\in$ \{VAR, PTR\}}{
        \If{Encoding is \texttt{Symbolic} or $A_i$ is not a \texttt{use} of a definition}{
            $\llbracket \mathbf{A_i} \rrbracket \leftarrow$ seedEmbeddings[``VAR'' or ``PTR'']
        }
        \Else{
            \For{each reaching definition RD of $A_i$}{
                $\llbracket \mathbf{RD} \rrbracket \leftarrow$ getInstrVec(RD)\;
                \label{algo:obtainEmb}
                \If{\hyperref[algo:obtainEmb]{$\llbracket \mathbf{RD} \rrbracket$} leads to cyclic dependency}{
                    Resolve and obtain $\llbracket \mathbf{RD} \rrbracket$  as shown in~\ref{subsubsec:ResolvingCirDep}
                }
                $\llbracket \mathbf{A_i} \rrbracket \leftarrow \llbracket \mathbf{A_i} \rrbracket + \llbracket \mathbf{RD} \rrbracket$
            }
        }
    }
    \ElseIf{$A_i\ $is a CONST}{
        $\llbracket \mathbf{A_i} \rrbracket  \leftarrow$  seedEmbeddings[``CONST'']\;
    }
    \ElseIf{$A_i\ $is a address of Basic block}{
        $\llbracket \mathbf{A_i} \rrbracket  \leftarrow$  seedEmbeddings[``LABEL'']\;
    }
    $\llbracket \mathbf{A} \rrbracket \leftarrow \llbracket \mathbf{A} \rrbracket\ + \llbracket \mathbf{A_i} \rrbracket$ 
}
\Return{$W_o * \llbracket \mathbf{O} \rrbracket\ + W_t * \llbracket \mathbf{T} \rrbracket\ + W_a * \llbracket \mathbf{A} \rrbracket $}
\caption{getInstrVec(Instruction I,  Dictionary seedEmbeddings)}
\end{procedure}
\begin{procedure}[H]
\label{algorithm:function-vec}
\DontPrintSemicolon
\SetAlgoNoLine
\footnotesize
\If{$\llbracket \mathbf{F} \rrbracket$ is already computed}{
\Return{$\llbracket \mathbf{F} \rrbracket$}}
$\llbracket \mathbf{F} \rrbracket \leftarrow \emptyset$\tcp*{Initializing n-d Function vector to zeroes}
\For{each basic block $BB_i$ $\in$ F}{
    $\llbracket \mathbf{BB_i} \rrbracket \leftarrow \emptyset$\tcp*{Initializing n-d Basic block vector to zeroes}
    \For{each \textit{live} instruction I $\in BB_i$}{
        $\llbracket \mathbf{I} \rrbracket \leftarrow$ getInstrVec(I,             seedEmbeddings)\;
        $\llbracket \mathbf{BB_i} \rrbracket \leftarrow \llbracket \mathbf{BB_i} \rrbracket  + \llbracket \mathbf{I} \rrbracket $\;
    }
    $\llbracket \mathbf{F} \rrbracket \leftarrow \llbracket \mathbf{F} \rrbracket  + \llbracket \mathbf{BB_i} \rrbracket $\;
}
\Return{$\llbracket \mathbf{F} \rrbracket$}
\caption{getFuncVec(Function F, Dictionary seedEmbeddings)}
\end{procedure}

The vector to represent a function $F$ with basic blocks $BB_1, BB_2,\ldots, BB_b$ is calculated as the sum of vectors of all its basic blocks as:
\begin{equation}
\llbracket \mathbf{F} \rrbracket = \sum_{i=1}^{b}\llbracket \mathbf{BB_i} \rrbracket
\end{equation}

Our encoding and propagation also take care of programs with function calls; the embeddings are obtained by using the call graph information. For every function call, the function vector for the callee function is calculated, and this value is used to represent the call instruction. For the functions that can be resolved only during link time, we just use the embeddings obtained for the \texttt{call} instruction. The final vector that is obtained encodes the function. 
The above procedure is applicable for recursive function calls as well.
This process of obtaining the instruction vector and function vector is outlined in Algorithms~\ref{algorithm:inst-vec} and~\ref{algorithm:function-vec}.

The procedure \texttt{getInstrVec} (Algorithm~\ref{algorithm:inst-vec}) computes the vector representation for a particular instruction $I$. The representations of the opcode ($\llbracket\mathbf{O}\rrbracket$) and type ($\llbracket\mathbf{T}\rrbracket$) of the instruction are fetched from the seed embedding vocabulary. 
For computing the representation of $\llbracket\mathbf{A}\rrbracket$, we iterate over the list of arguments of the instruction. If the argument corresponds to a function and the definition of that function is available, we find the representation of that function and use it as the representation of the argument. If the definition of the function is not available, then the generic representation of function in the seed embedding vocabulary is used to represent the argument.

If the argument is a variable or pointer, we compute the representation of all its reaching definitions and sum them up to represent the argument, in case of flow aware encodings. For symbolic encodings, we use a simpler procedure by using the generic representation of variable or pointer from the seed embedding vocabulary as the representation of the argument.

For the other cases, when the argument is a constant or an address of a basic block (its label), the corresponding generic representations are used.
Finally, representations of all the arguments (of the instruction) are summed up to compute $\llbracket\mathbf{A}\rrbracket$. And, the instruction vector is computed as the weighted sum of $\llbracket\mathbf{O}\rrbracket$, $\llbracket\mathbf{T}\rrbracket$ and $\llbracket\mathbf{A}\rrbracket$.

The procedure \texttt{getFuncVec} (Algorithm~\ref{algorithm:function-vec}) computes the function vector. First, the representation of every basic block in the function is calculated. The instruction vectors corresponding to the live instructions are calculated by making a call to the procedure \texttt{getInstrVec}
and the basic block vector is obtained as the element-wise sum of these instruction vectors. Which, when summed up, forms the function vector.

If $\llbracket \mathbf{F_1} \rrbracket, \llbracket \mathbf{F_2} \rrbracket, \ldots , \llbracket \mathbf{F_f} \rrbracket$ are the embeddings of the functions $F_1, F_2, \ldots , F_f$ in a program, then the code vector representing the program $P$ is calculated as the sum of the embeddings of all such functions as:
\begin{equation}
\llbracket \mathbf{P} \rrbracket = \sum_{i=1}^{f}\llbracket \mathbf{F_i} \rrbracket
\end{equation}
\section{Experimental Results}
\label{sec:experimentation}

We used the SPEC CPU 17~\cite{spec17-Bucek:2018:SCN:3185768.3185771} benchmarks and Boost library~\cite{boost} as the datasets for learning the representations. 
The programs in these benchmarks are converted to LLVM IR by varying the compiler optimization levels (\texttt{-O[0--3], -Os, -Oz}) at random.
The $\langle h, r, t \rangle$ triplets are created from the resultant IRs using the relations that were explained in Sec.~\ref{subsec:preprocessing}. Embeddings of such triplets are learned using an open-source implementation of TransE~\cite{han2018openke}, which was explained in Sec. ~\ref{subsec:rep-learning}.

We wrote an LLVM analysis pass to extract the generic tuples from the IR so that they can be mapped to form triplets. There were $\approx134M$ triplets in the dataset, out of which $\approx8K$ relations were unique; from these, we obtain 64 different entities whose embeddings were learnt. The training was done with SGD optimizer for 1500 epochs to obtain embedding vectors of 300 dimensions. 

These learned embeddings of 64 different entities form the \textit{seed embeddings}. 
Additional related information is listed in the Tab.~\ref{tab:comparisonMatrix}. 
We heuristically set $W_o$, $W_t$ and $W_a$ to 1, 0.5, 0.2, respectively. The vectors at various levels were computed using another LLVM pass.

In this section, we attempt to answer \hyperref[RQ1]{\textbf{RQ1}} by showing the clusters formed by the entities of \textit{seed embedding vocabulary} and  \hyperref[RQ2]{\textbf{RQ2}} by showing the effectiveness of \irtovec embeddings on two different optimization tasks.

\subsection{RQ1: Evaluation of Seed Embeddings}
\label{subsection:seedEmbeddings-eval}
The seed embeddings are analyzed to demonstrate whether the semantic relationship between the entities are effectively captured by forming clusters.
These clusters show that \irtovec is able to capture the characteristics of the instructions and the arguments.

The entities corresponding to the obtained seed embeddings 
are categorized as groups based on the nature of the operations they perform --- \textit{Arithmetic operators} containing the integer and floating-point type arithmetic operations, \textit{Pointer operators} which access memory, \textit{Logical operators} which perform logical operations, \textit{Terminator operators} which form the last instruction of the basic block, \textit{Casting operators} that perform typecasting, \textit{Type} information and \textit{Arguments}.

\begin{figure}
    \captionsetup[subfigure]{font=scriptsize,labelfont=scriptsize}
    \centering
    \subfloat[\scriptsize{Types}]{{\includegraphics[scale=0.3]{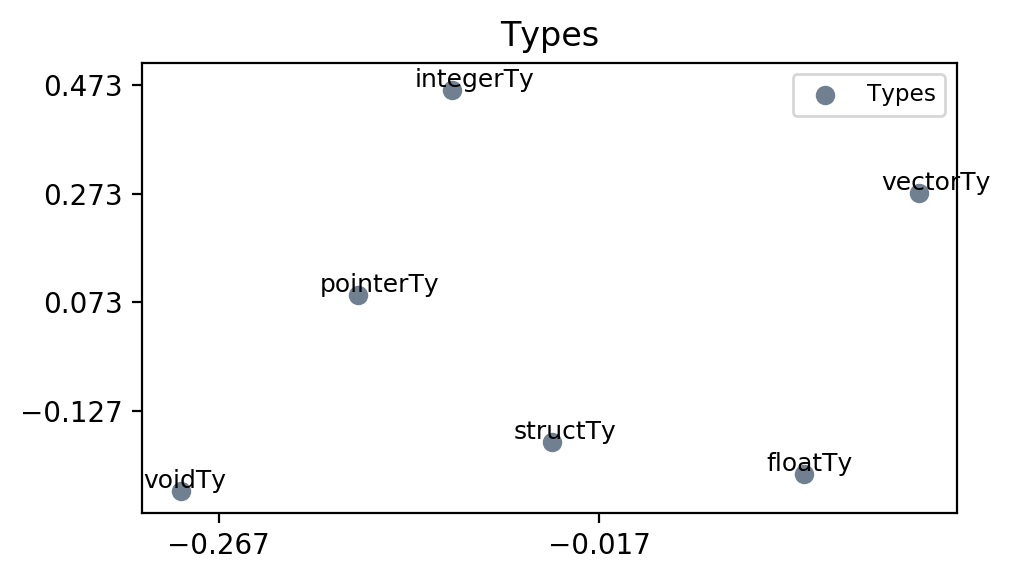} }}%
    \qquad
    \subfloat[\scriptsize{Arithmetic Operations Vs. Arguments}]{{\includegraphics[scale=0.3]{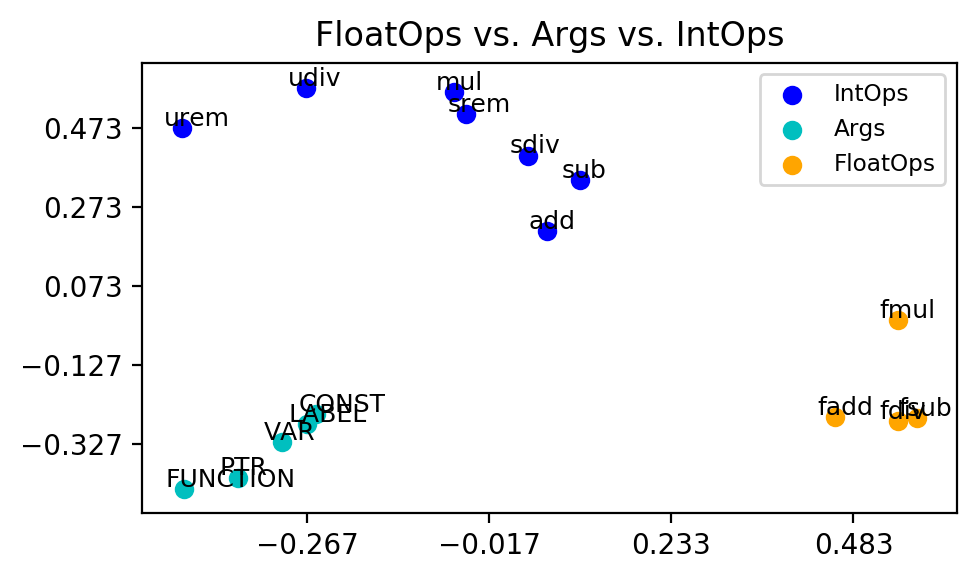} }}%
    \qquad
    \subfloat[\scriptsize{Arguments Vs. Logical Operations}]{{\includegraphics[scale=0.3]{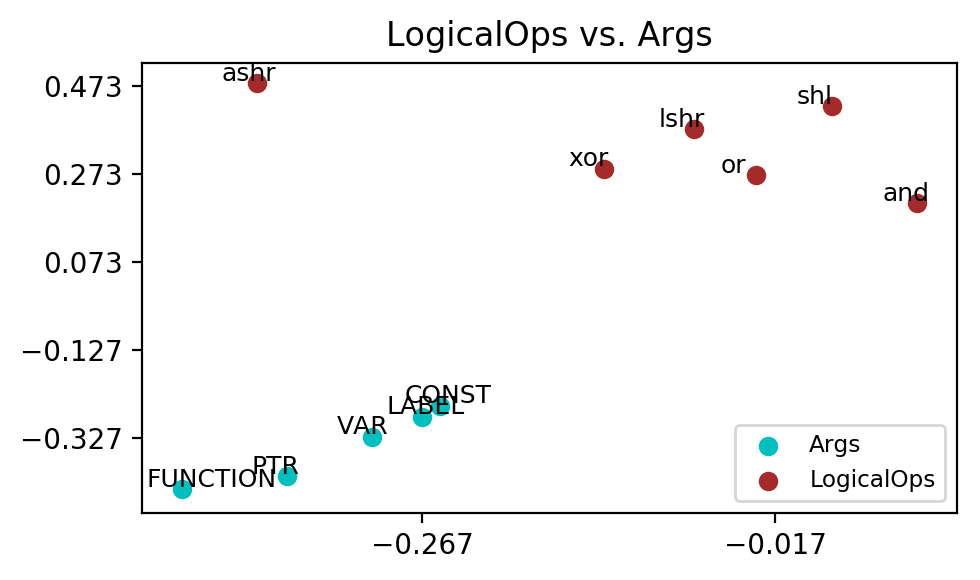} }}%
    \qquad
    \subfloat[\scriptsize{Arithmetic Vs. Casting Operations}]{{\includegraphics[scale=0.3]{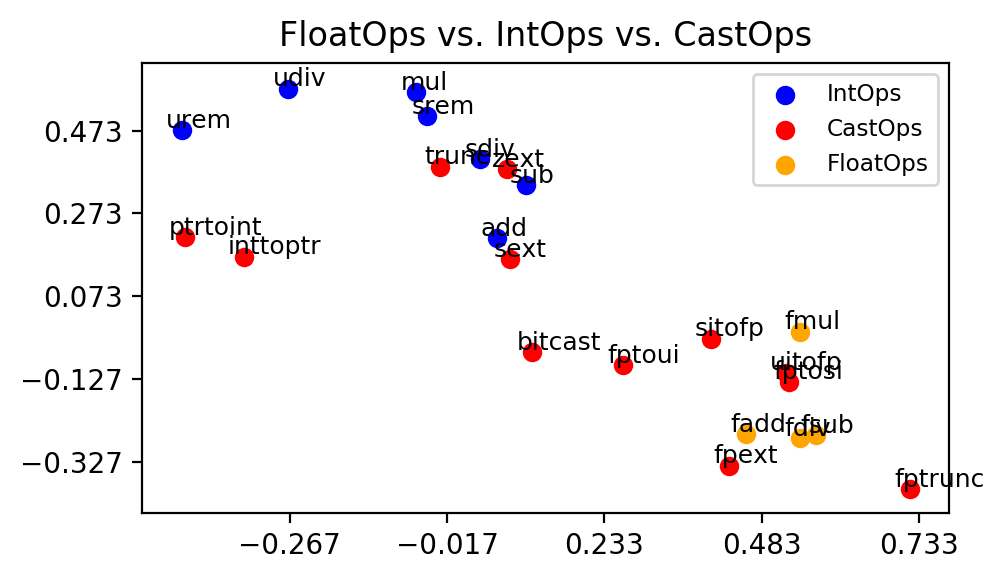} }}%
    \qquad
    \subfloat[\scriptsize{Pointer Vs. Arithmetic Operations}]{{\includegraphics[scale=0.3]{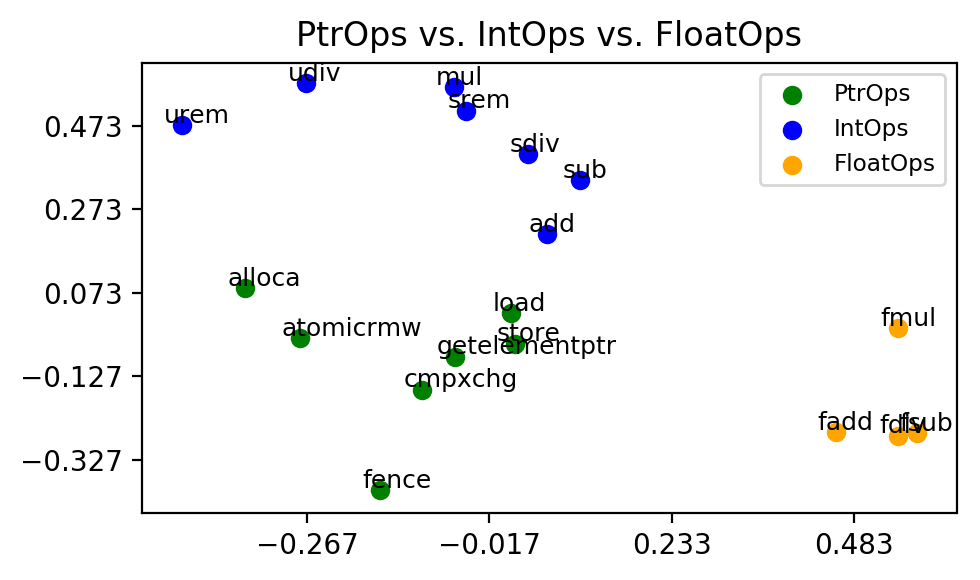} }}%
    \qquad
    \subfloat[\scriptsize{Terminator Vs. Logical Vs. Arithmetic Operations}]{{\includegraphics[scale=0.3]{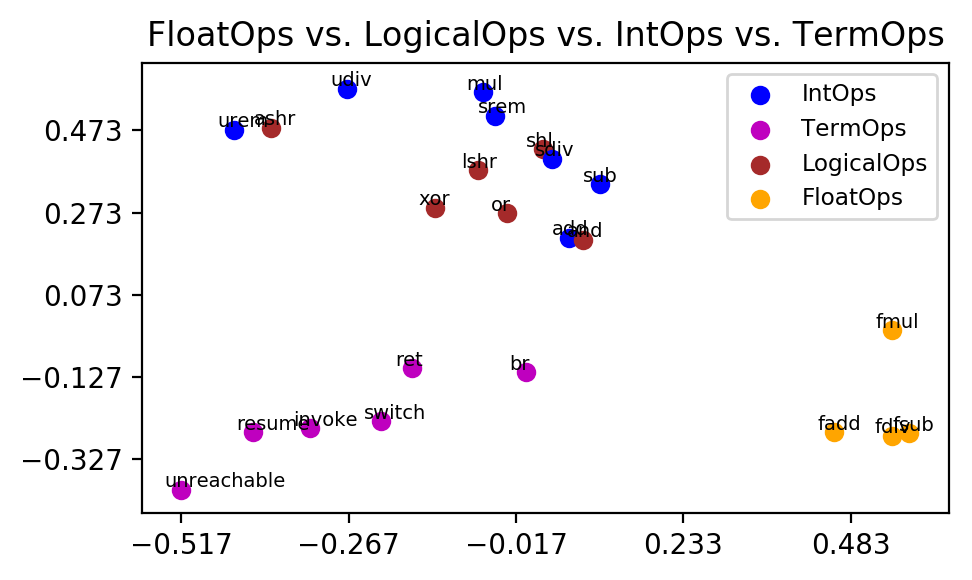} }}%
    \caption{Comparison of embeddings of various seed entities}
    \vspace*{-\baselineskip}
    \label{fig:clusters}
\end{figure}

Clusters showing these groups are plotted using PCA (Principal Component Analysis), a dimensionality reduction mechanism to project the points from the higher to lower dimensional space~\cite{wold1987principalcomponentanalysis}. Here, to visualize the 300-dimensional data in 2 dimensions, we use 2-PCA, and the resulting clusters are shown in Fig.~\ref{fig:clusters}.

In Fig.~\ref{fig:clusters}(a), we show the relation between various types. It can be observed that \texttt{vectorTy} being an aggregate type can accept any of the other first-class types and lies approximately equidistant from \texttt{integerTy}, \texttt{pointerTy}, \texttt{structTy} and \texttt{floatTy}. And, \texttt{vectorTy} lies farthest from \texttt{voidTy} justifying that it is unlikely that \texttt{voidTy} elements will be aggregated together as vectors. 

In Fig.~\ref{fig:clusters}(b), we show that all integer based arithmetic operators are grouped together and are distinctly separated from floating-point based operators.
It can be said that the analogies between the operators are captured. 
For example, the distance between (\texttt{add}, \texttt{fadd}) is similar to that of the distance between (\texttt{sub}, \texttt{fsub}).
From Fig.~\ref{fig:clusters}(b), ~\ref{fig:clusters}(e) and ~\ref{fig:clusters}(f), it can also be seen that the arithmetic operators are distinctly separated from the arguments, pointer operators and terminator operators. 
Similarly, from Fig.~\ref{fig:clusters}(c), we can see that the logical operators are also distinctly separated from the arguments.

In Fig.~\ref{fig:clusters}(d), we show the relationship between arithmetic and casting operators. It can be clearly seen that the integer based casting operators like \texttt{trunc, zext, sext }, etc. are grouped together with integer operators, and the floating-point based casting operators like \texttt{fptrunc, fpext, fptoui}, etc. are grouped together with floating-point operators.
On observing Fig.~\ref{fig:clusters}(d) and Fig.~\ref{fig:clusters}(e), it can be seen that \texttt{ptrtoint} and \texttt{inttotpr} are closer to both integer operators and pointer operators. Fig.~\ref{fig:clusters}(e) also demonstrates that the arithmetic operators are clearly distinct from pointer operators.
Logical operators operate on integers, and hence they are grouped together with integer operators, as observed in Fig.~\ref{fig:clusters}(f). 

In summary, these clusters show that the obtained seed embeddings are indeed meaningful as they capture intrinsic syntactic and semantic relationships of LLVM entities.

\subsection{RQ2 (a): Heterogeneous Device Mapping}
\label{subsection:dev-map}
Grewe et al.~\cite{o2013portable-grewe} proposed the heterogeneous device mapping task to map OpenCL kernels to the optimal target device - CPU or GPU in a heterogeneous system.
In this experiment, we use the embeddings obtained by \irtovec to map OpenCL kernels to its optimal target.

\subsubsection*{Dataset} 
We use the dataset provided by Ben-Nun et al.~\cite{ncc} for this experiment. It consists of 256 unique OpenCL kernels from seven different benchmark suites comprising of AMD SDK, NPB, NVIDIA SDK, Parboil, Polybench, Rodinia and SHOC. 
Taking the kernels and varying their two auxiliary inputs (data size and workgroup size) for each kernel, a dataset is obtained. This leads to about 670 CPU or GPU labelled data points for each of the two GPU devices -- AMD Tahiti 7970 and NVIDIA 970. 

\subsubsection*{Experimental Setup}
The embeddings for each kernel are computed using \irtovec infrastructure. Gradient boosting classifier with a learning rate of 0.5, which allows a maximum depth of up to 10 levels and 70 estimators with ten-fold cross-validation is used to train the model.

We use simpler models like gradient boosting classifier, as our embeddings are generated at the program/function level directly without forming sequential data that need sequential neural networks like RNNs or LSTMs.
More advantages of our modelling are discussed in Sec.~\ref{subsection:trainingchar}. Similar to the earlier methods, we use the runtimes corresponding to the predicted device, to calculate the speedup against the static mapping heuristics proposed by Grewe et al.~\cite{o2013portable-grewe}.

We compare the prediction accuracy and speedup obtained by \irtovec across two platforms (AMD Tahiti 7970 and NVIDIA GTX 970) with the manual feature-extraction approach of Grewe et al.~\cite{o2013portable-grewe} and the state-of-the-art methodologies of DeepTune~\cite{cummins2017end2end}, and inst2vec~\cite{ncc}. 

\subsubsection*{Accuracy}
\begin{figure}
    \centering
    \includegraphics[scale=0.7]{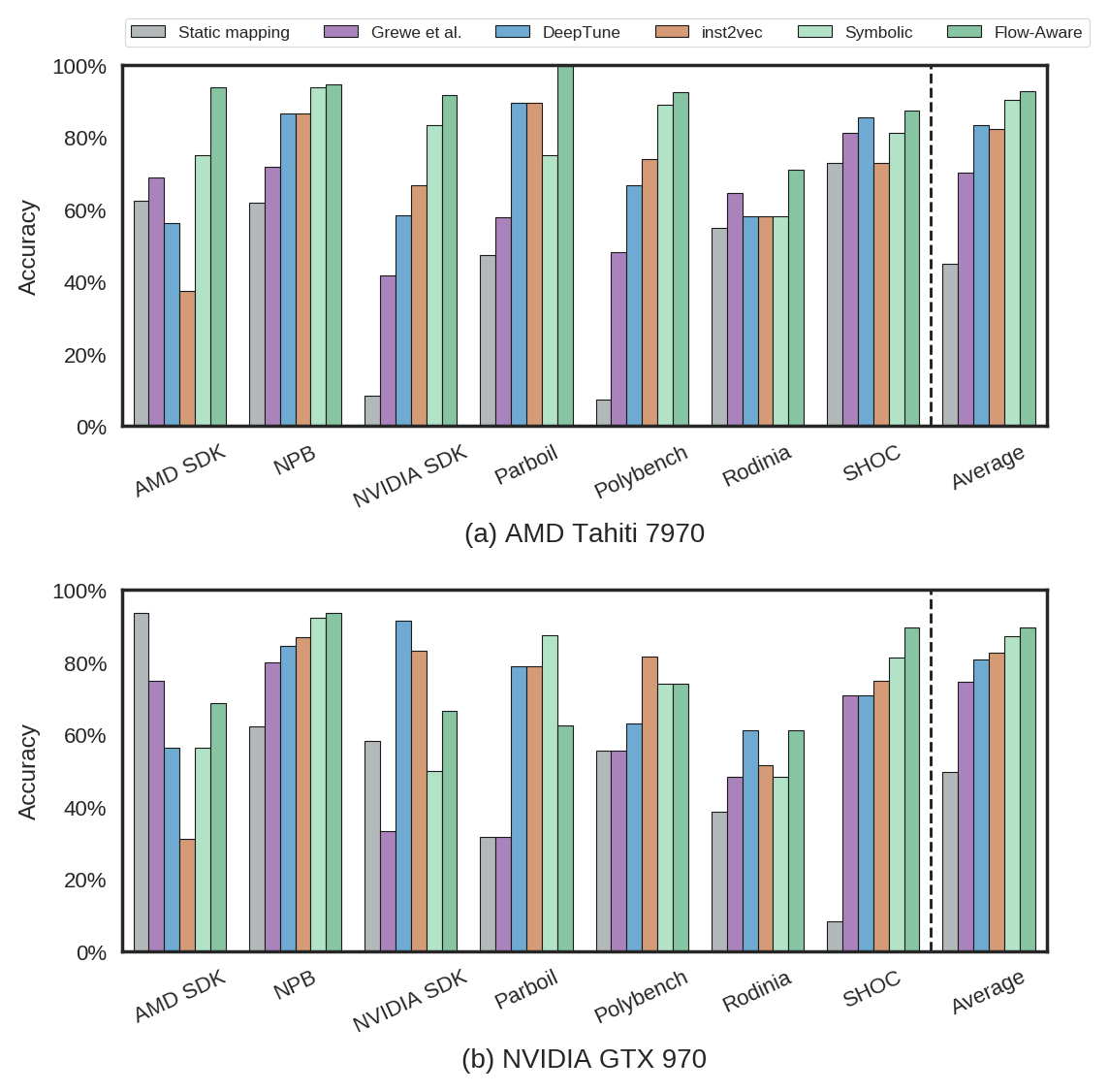}
    \vspace{-\baselineskip}
     \caption{Accuracy for heterogeneous device mapping task on various benchmark suites. 
    }
    \label{fig:devmapAccuracy}
\end{figure}

Accuracy is computed as the percentage of correct device mapping predictions for a kernel by the model over the total number of predictions during test time.
In Fig.~\ref{fig:devmapAccuracy}, we show the accuracy of mapping using the encodings generated by \irtovec and other methods. \textit{Flow-Aware} and \textit{Symbolic} encodings achieve an average accuracy of 91.26\% and 88.72\% accuracy, respectively. 

In Tab.~\ref{tab:predictionAccuracy_dm}, we show the percentage improvement of accuracy obtained by \textit{Flow-Aware} encodings over the other methods. It can be observed that the \textit{Flow-Aware} encodings achieve higher performance over the other methods~\cite{o2013portable-grewe, cummins2017end2end, ncc}. 
On AMD Tahiti 7970, our \textit{Flow-Aware} encodings achieve the highest accuracy in all 7 benchmark suites. In addition, our \textit{Symbolic} encodings achieve second-highest in 4/7 benchmark-suites. Similarly, in NVIDIA GTX 970, \textit{Flow-Aware} encodings achieve the highest accuracy in 3/7 cases, and \textit{Symbolic} encodings achieve the highest or second-highest accuracy in 4/7 cases. 

Along with inst2vec encodings, the NCC framework~\cite{ncc} also proposes the inst2vec-imm encodings to handle immediate values. For this, they formulate four \textit{immediate value handling methods}. As there is no single consistent winner among these value handling methods, NCC considers the best result out of them.
As the benchmark-wise results are not published, we are unable to compare our results with inst2vec-imm in detail.

\begin{table}[h]
  \caption{\% improvement in accuracy obtained by Flow-Aware encodings when compared to the other methods}
  \vspace*{-\baselineskip}
  \label{tab:predictionAccuracy_dm}    \small
  \begin{tabular}{p{2.5cm}p{1.9cm}p{1.6cm}p{1.5cm}p{2.2cm}p{1.5cm}p{1.5cm}}
    \toprule
    \textbf{Architecture} & \textbf{Grewe et al. \cite{o2013portable-grewe}} & \textbf{DeepTune \cite{cummins2017end2end}} & \textbf{inst2vec \cite{ncc}} & \textbf{inst2vec-imm\footnotemark \cite{ncc}}& \textbf{\irtovec  Symbolic}\\
    \midrule
    \textbf{AMD Tahiti 7970} & 26.50\% & 10.93\% & 12.12\% & 5.38\% & 2.81\% \\
        \textbf{NVIDIA GTX 970} & 22.96\% & 11.70\% & 9.69\% & 3.54\% & 2.92\% \\
\bottomrule
\end{tabular}
\end{table}
\footnotetext{The numbers are quoted from the paper.}

\begin{figure}
  \centering 
    \begin{tabular}{@{}c@{}}
        \includegraphics[width=\textwidth]{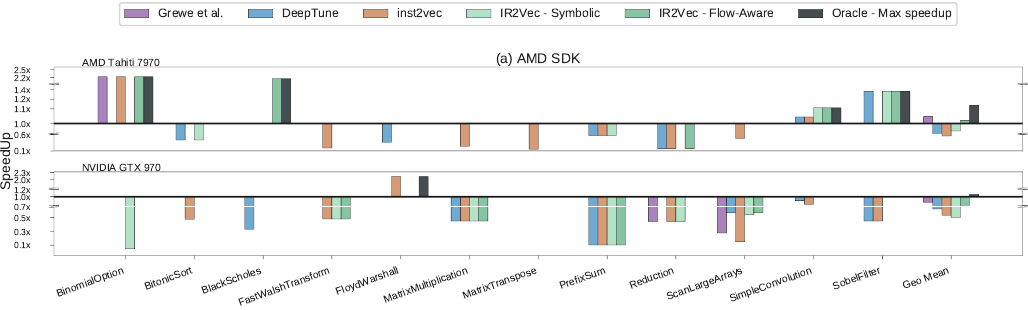}\\
        \includegraphics[width=\textwidth]{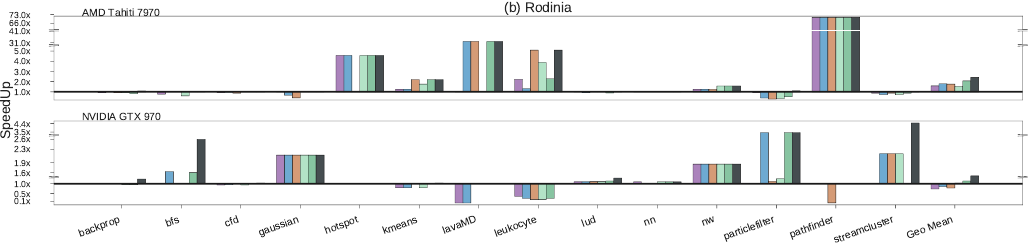}\\
        \includegraphics[width=\textwidth]{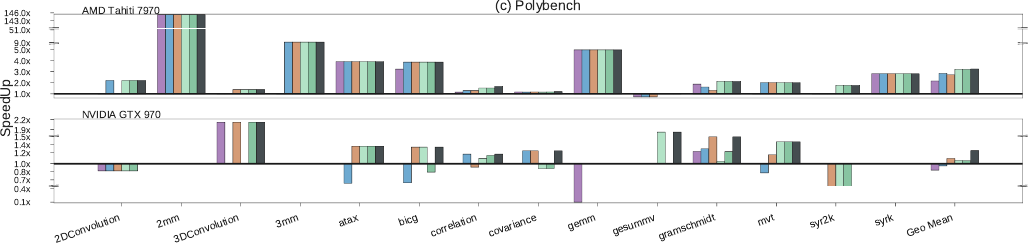}\\
        \includegraphics[width=\textwidth]{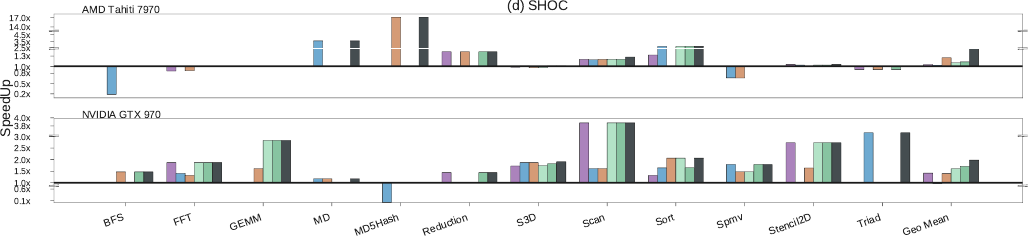}\\
    \end{tabular}
  \caption{Plot showing the speedups corresponding to the device mapping predictions by various methods}
  \label{fig:dmSpeedup}
\end{figure}

\begin{figure}
  \ContinuedFloat 
  \centering 
    \begin{tabular}{cc}
        \includegraphics[width=0.48\textwidth]{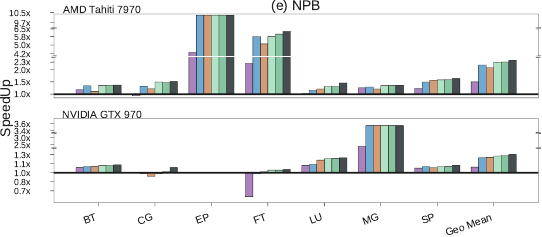} &
        \includegraphics[width=0.48\textwidth]{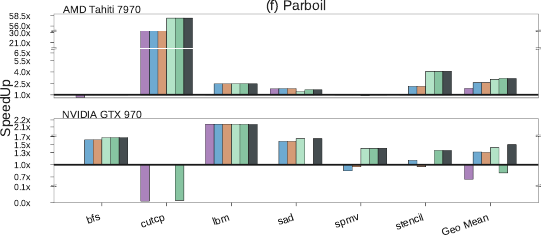}\\
        \vtop{%
          \vskip-2.5ex 
          \hbox{%
            \includegraphics[width=0.48\textwidth]{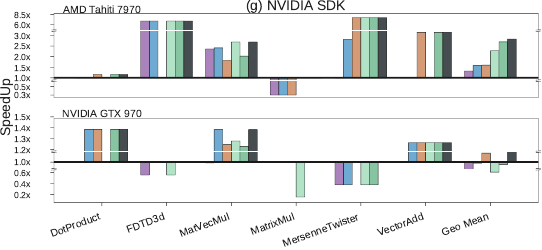}
          }%
        } 
        
        &
        \vtop{%
          \vskip-2.5ex 
          \hbox{%
            \includegraphics[width=0.48\textwidth]{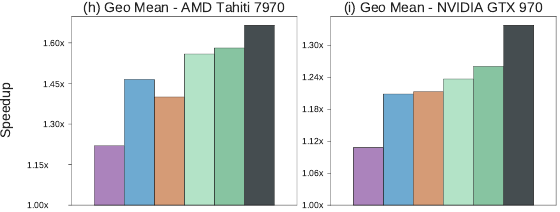}
          }%
        }
    \end{tabular}
  \caption{\textit{(Contd.)} Plot showing the speedups corresponding to the device mapping predictions by various methods}
\end{figure}

\textit{Speedups.}
In Fig.~\ref{fig:dmSpeedup},
we show the speedups achieved in comparison with static mapping as the baseline on both the platforms under consideration. 
With \textit{Flow-Aware} encodings, we achieve about 88.38\% and 77.65\% of the maximum speedup given by the oracle on AMD Tahiti and NVIDIA GTX resepectively, when compared to the 70.76\% on AMD Tahiti, and 70.66\% on NVIDIA GTX by inst2vec and 72.90\% on AMD Tahiti and 66.79\% on NVIDIA GTX by Deeptune. (With \textit{Symbolic} encodings we achieve about 80.11\% and 72.03\% of the maximum speedup on AMD Tahiti and NVIDIA GTX.)

\textit{Flow-Aware} encodings achieve a geometric mean of $ 1.58\times$ speedup on AMD Tahiti 7970 and a $ 1.26\times$ speedup on NVIDIA GTX 970; in comparison, on the AMD platform, the speedups achieved by the state-of-the-art models of DeepTune~\cite{cummins2017end2end} and inst2vec~\cite{ncc} is $ 1.46\times$ and $ 1.4\times$ respectively; while, on the NVIDIA platform, both DeepTune and inst2vec achieve a comparable $ 1.21\times$ speedups.
\textit{Symbolic} encoding also performs better than the state-of-the-art by achieving a speedup of $ 1.56\times$ on AMD Tahiti and $ 1.24\times$ on NVIDIA GTX. 

\begin{table}[h]
  \caption{\% improvement in speedup obtained by Flow-Aware encodings when compared to the other methods}
  \vspace*{-\baselineskip}
  \label{tab:predictionSpeedup_dm}    \small
  \begin{tabular}{p{2.5cm}p{1.9cm}p{1.6cm}p{1.5cm}p{2.2cm}}
    \toprule
    \textbf{Architecture} & \textbf{Grewe et al. \cite{o2013portable-grewe}} & \textbf{DeepTune \cite{cummins2017end2end}} & \textbf{inst2vec \cite{ncc}}\footnotemark & \textbf{\irtovec \hfill\break Symbolic}\\
    \midrule
    \textbf{AMD Tahiti 7970} & 29.65\% & 7.96\% & 12.95\% & 1.43\% \\
    \textbf{NVIDIA GTX 970} & 13.77\% & 4.31\% & 3.93\% & 1.92\% \\
\bottomrule
\end{tabular}
\end{table}
\footnotetext{The inst2vec-imm results given in their paper~\cite{ncc} are based on arithmetic mean, whereas our results are based on geometric mean. inst2vec-imm is reported to achieve a (arithmetic mean) speedup of $ 3.47\times$ and $ 1.44\times$; whereas our \textit{Flow-Aware} encodings achieve a speedup of $ 3.51\times$ and $ 1.47\times$ on AMD Tahiti and on NVIDIA GTX respectively.
}

In Tab.~\ref{tab:predictionSpeedup_dm}, we show the percentage improvement of speedups obtained by \textit{Flow-Aware} encodings in comparison with the earlier methods. It can be seen that the \textit{Flow-Aware} encodings perform better on both the platforms when compared to the other works.

On benchmarks like \texttt{ep} (\textit{embarrassingly parallel}), where mapping them to GPUs gets the optimal performance~\cite{o2013portable-grewe}, we map to GPU---in all the turns of 10 fold cross-validation---and hence achieve the highest possible speedup. Similarly, for \texttt{LU}, which gives an optimal performance when mapped to CPUs~\cite{o2013portable-grewe}, we map the kernels that ought to be mapped to CPU correctly, with very high confidence (of about 95\%) in various turns of 10 fold cross-validation. In comparison, DeepTune and inst2vec map LU to CPUs with a confidence of 77\% and 85\%, respectively.

\textit{Slowdown.} Our predictions using \textit{Flow-Aware} encodings result in the \textit{least} number of slowdowns both at the level of benchmarks and benchmark-suites.  Our predictions result in a slowdown in 18/142 of benchmarks across two platforms; in comparison, inst2vec and DeepTune result in a slowdown in 33 and 32 benchmarks, respectively. 
Also, at an aggregate benchmark-suite level, the prediction using \textit{Flow-Aware} encodings leads to a slowdown in 3/14 cases across platforms, in comparison inst2vec and DeepTune result in slowdowns in three and six cases, respectively.

\subsection{RQ2 (b): Prediction of optimal Thread Coarsening factor}
\label{subsection:thread-coarsening}
Thread coarsening~\cite{Volkov-Demmel-10.5555/1413370.1413402} is the process of increasing the work done by a single thread by fusing two or more concurrent threads. Thread coarsening factor corresponds to the number of threads that can be fused together. Selection of an optimal thread coarsening factor can lead to significant improvements~\cite{Magni-SC13-DBLP:conf/sc/MagniDO13} in the speedups on GPU devices and a naive coarsening would lead to a substantial slowdown.

A thread coarsening factor of a kernel that gives the best speedup on a GPU could give the worst performance with the same coarsening factor on another device (either within or across vendors) because of the architectural characteristics of the device~\cite{magni2014automatic, Stawinoga-10.1145/3194242}. For example, \texttt{nbody} kernel, which has a higher degree of Instruction Level Parallelism, can be better exploited by VLIW based AMD Radeon than SIMD based AMD Tahiti~\cite{magni2014automatic}.

\subsubsection*{Dataset} 
In this experiment, we follow the experimental setup proposed by Magni et al.~\cite{magni2014automatic} to predict the optimal thread coarsening factor---among $\{1,2,4,8,16,32\}$---for a given kernel specific to a GPU device. Even for this experiment, we use the dataset provided by Ben-Nun et al.~\cite{ncc}. It consists of about 68 datapoints from 17 OpenCL kernels on 4 different GPUs -- AMD Radeon 5900, AMD Tahiti 7970, NVIDIA GTX 480 and NVIDIA Tesla K20c.
These kernels are collectively taken from AMD SDK, NVIDIA SDK and Parboil benchmarks.
A datapoint consists of the kernel and its runtime corresponding to each thread coarsening factor on a particular GPU device.

\subsubsection*{Experimental Setup}
Even for this task, we use gradient boosting classifier instead of LSTMs and RNNs to predict the coarsening factor for the four GPU targets. For this experiment, we set the learning rate as 0.05 with 140 decision stumps with 1 level, as the number of data points in the dataset is very low. We use ten-fold cross-validation for measuring the performance.

\begin{table}[h]
  \caption{\% improvement in speedup obtained by Flow-Aware encodings when compared to the other methods}
  \vspace*{-\baselineskip}
  \label{tab:predictionSpeedup-tc}    \small
  \begin{tabular}{p{2.5cm}p{1.9cm}p{1.6cm}p{1.5cm}p{2.2cm}p{1.5cm}}
    \toprule
    \textbf{Architecture} & \textbf{Magni et al. \cite{o2013portable-grewe}} & \textbf{DeepTune \cite{cummins2017end2end}} & \textbf{DeepTune-TL \cite{cummins2017end2end}} & \textbf{inst2vec \cite{ncc}}\footnotemark  & \textbf{\irtovec  Symbolic}\\
    \midrule
    \textbf{AMD Radeon 5900} & 27.66\% & 5.26\% & 5.26\% & 4.35\% & -- \\
    \textbf{AMD Tahiti 7970} & 25.41\% & 29.37\% & 36.56\% & 18.17\% & 2.08\% \\
    \textbf{NVIDIA GTX 480} & 45.31\% & 25.21\% & 18.89\% & 23.89\% & 3.98\%\\
    \textbf{NVIDIA Tesla K20c} & 52.84\% & 15.41\% & 11.98\% & 11.98\% & 0.18\%\\
\bottomrule
\end{tabular}
\end{table}
\footnotetext{As per the results given in the NCC paper~\cite{ncc}, inst2vec-imm achieves a (arithmetic) mean of $ 1.28\times$, $ 1.18\times$, $ 1.11\times$ and $1\times$ speedup on AMD Radeon, AMD Tahiti, NVIDIA GTX and NVIDIA Tesla; whereas our \textit{Flow-Aware} encodings achieve a (arithmetic) mean of $ 1.25\times$, $ 1.3\times$, $ 1.26\times$ and $ 1.16\times$ respectively.}

\begin{figure}
    \centering
    \begin{tabular}{@{}c@{}}
        \includegraphics[scale=0.55, width=\textwidth]{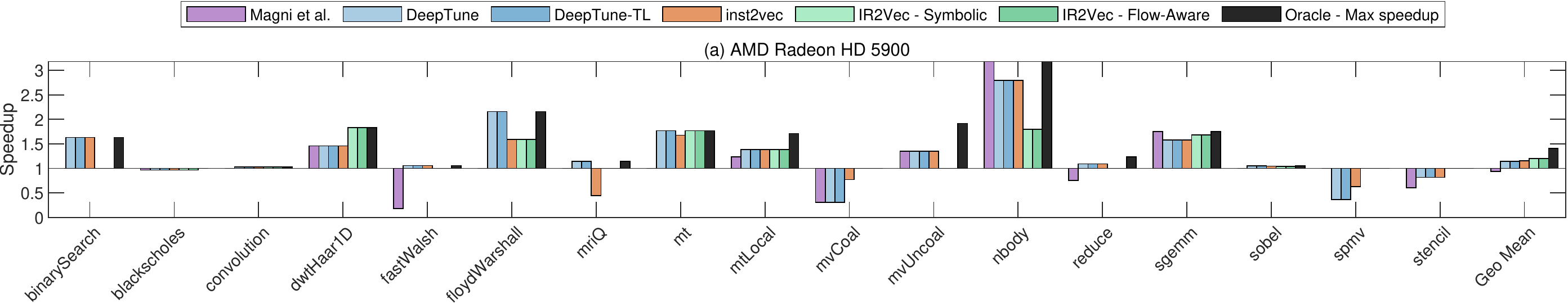}
    \end{tabular}
    \begin{tabular}{@{}c@{}}
        \includegraphics[scale=0.55, width=\textwidth]{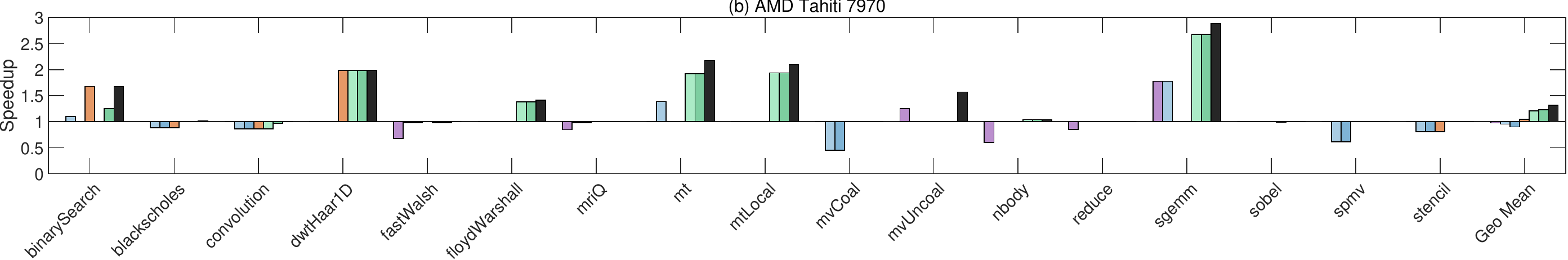}
    \end{tabular}
    \begin{tabular}{@{}c@{}}
        \includegraphics[scale=0.55, width=\textwidth]{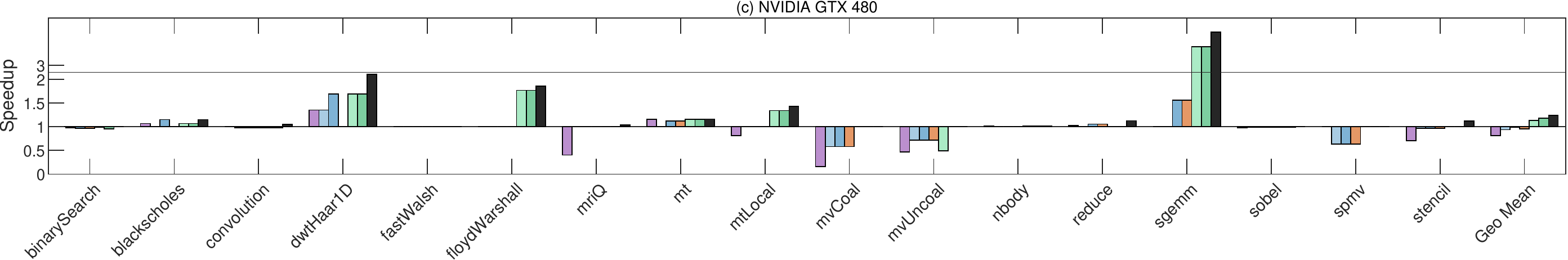}
    \end{tabular}
    \begin{tabular}{@{}c@{}}
        \includegraphics[scale=0.55, width=\textwidth]{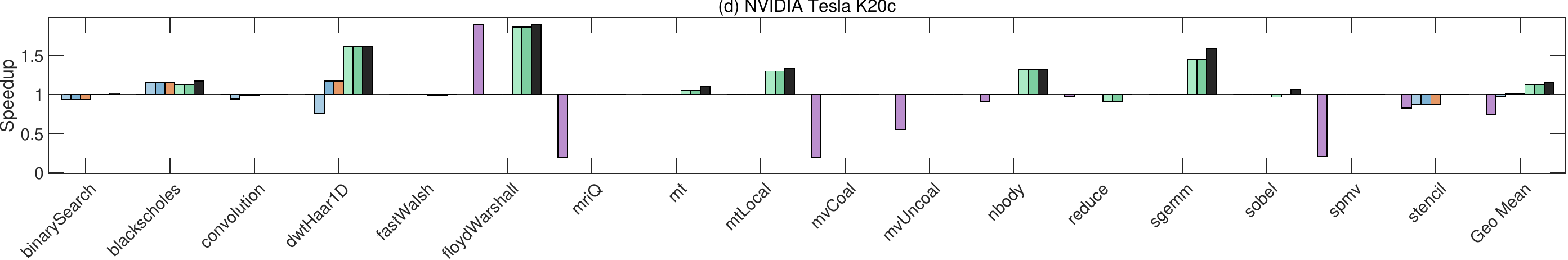}
    \end{tabular}
    \caption{Plot showing the speedups achieved by predicted coarsening factors by various methods}
    \label{fig:tcSpeedup}
     \vspace*{-0.4cm}
\end{figure}

\textit{Speedups.}
In Fig.~\ref{fig:tcSpeedup}, we show the speedups achieved by our encodings and earlier works on four different platforms -- AMD Radeon HD 5900, AMD Tahiti 7970, NVIDIA GTX 480 and NVIDIA Tesla K20c. 

On AMD Radeon, both of our encodings achieve a speedup of $ 1.2\times$ when compared to the state-of-the-art speedup of $ 1.15\times$ and $ 1.14\times$ achieved by inst2vec~\cite{ncc} and DeepTune model with transfer learning (DeepTune-TL)~\cite{cummins2017end2end}.
In AMD Tahiti, \textit{Flow-Aware} encodings achieve a speedup of $ 1.23\times$; \textit{Symbolic} encoding achieves a speedup of $ 1.2\times$, whereas the earlier works by DeepTune-TL and inst2vec achieve a speedup of $ 0.9\times$ and $ 1.04\times$ respectively. In Tab.~\ref{tab:predictionSpeedup-tc}, we show the percentage improvement of speedup obtained by \textit{Flow-Aware} encodings over other methodologies. From the table, we can infer that \textit{Flow-Aware} gives better results for every architecture when compared to the other methods. 

We are the \textit{first ones} to achieve a positive speedup on NVIDIA GTX 480; our \textit{Flow-Aware} and \textit{Symbolic} encodings obtain a speedup of $ 1.18\times$ and $ 1.13\times$ respectively when compared to $ 0.95\times$ and $ 0.99\times$ speedup achieved by inst2vec and DeepTune-TL. 
We get a speedup of $ 1.13\times$ with both of our proposed encodings on NVIDIA Tesla K20c. In contrast, DeepTune-TL and inst2vec obtain a speedup of $ 1.01\times$ on this platform.

On average, it can be seen that both encodings outperform the earlier methods for prediction of the thread coarsening factor on all the four platforms under consideration. 

\textit{Slowdown.}  Magni et al.~\cite{magni2014automatic} observe that \texttt{spmv} and \texttt{mvCoal} kernels have irregular dependences that causes a poor response to coarsening, and hence no performance improvement for them is possible.
For these kernels, \irtovec obtains the baseline speedup \textit{without} resulting in a slowdown. In contrast, the earlier models result in negative speedups (Deeptune results in a slowdown upto $0.36$ $\times$ in AMD Radeon and inst2vec results in a slowdown of upto $0.63$ $\times$ in AMD Radeon and NVIDIA GTX). 
The same argument applies for \texttt{stencil} kernel (an iterative Jacobi stencil on 3D-grid), where the coarsening leads to slowdown (except in NVIDIA GTX), while \irtovec still obtain the baseline speedup.

When compared to the other methods, we obtain the best speedup on about 70\% of the kernels on all platforms. 
It can be observed that the \textit{Flow-Aware} encodings \textit{rarely lead to slowdowns}; this happens in only 8/68 cases (17 benchmark-suits, across 4 platforms), even on these eight cases, the speedup is still close---within 10\%---of the baseline. 
Whereas, predictions by inst2vec and DeepTune-TL result in a slowdown in 18 and 21 cases.
We believe that this is because of the flow information associated with the obtained vectors. 
\section{RQ3: \irtovec - Perspectives}
\label{subsection:scalability}

We discuss some perspectives on \irtovec.  We answer \hyperref[RQ3]{\textbf{RQ3}} by doing a scalability study.
\vspace{-0.25\baselineskip}
\subsection{Training characteristics}
\label{subsection:trainingchar}
By design, training with \irtovec embeddings takes lesser training time. 
This is because our framework has the flexibility to model the embeddings as non-sequential data at program or function level. 
Whereas, other methods are limited to modelling the input programs only as sequential data, and hence are bound to using sequential models like LSTMs and RNNs. Using such models will involve training more number of parameters than non-sequential models like Gradient Boosting.

\paragraph{Device Mapping}
Training for the device mapping task by \irtovec takes $\approx 5$ seconds on a P100 GPU, when compared to about $10$ hours and $12$ hours of training time taken by DeepTune~\cite{cummins2017end2end} and NCC~\cite{ncc} respectively. This results in a \textit{reduction of about $\approx 7200\times$--$8640\times$ in training time without a reduction in performance}. 

The earlier works take much time for training because they involve training a large number of parameters: DeepTune~\cite{cummins2017end2end} uses  $\approx 77K$ parameters, while NCC~\cite{ncc} uses $\approx69K$ parameters. In contrast, the \irtovec predictions use Gradient Boosting, which is a collection of a small number of shallow decision trees.
This reduction in time is primarily possible because the embeddings obtained by \irtovec enable us effectively use the Gradient boosting algorithm instead of the compute-intensive and data-hungry neural networks (LSTM in this case) that do not fit well in the cache hierarchy. 

\paragraph{Thread Coarsening}
Even for the thread coarsening task, our model takes lesser time of $\approx$10 seconds for training when compared to $\approx$11 hours of training time needed by DeepTune-TL and $\approx$1 hour of training time needed (and 77K and 69K parameters used) by DeepTune and NCC approaches. 
This results in $\approx 360\times$--$3960\times$ reduction of training time, and again, achieving good speedups. 

\subsection{Symbolic vs. Flow-Aware}
\label{subsection:symbolicVsFlowaware}

\begin{figure}
    \centering
    \includegraphics[scale=0.34, width=\textwidth]{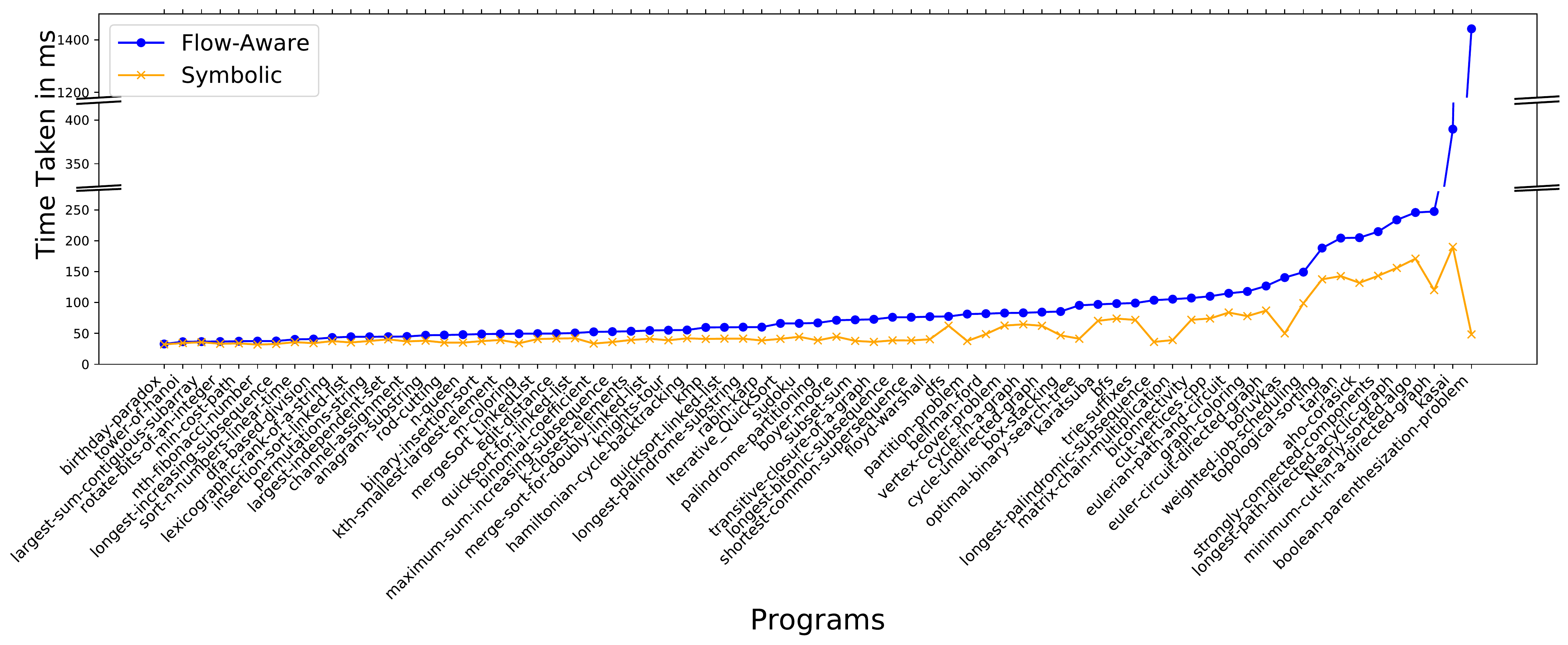}
    \caption{Comparison of time taken to generate the Symbolic and Flow-Aware encodings from \textit{Seed Embedding Vocabulary}}
    \label{fig:timetaken}
     \vspace*{-0.4cm}
\end{figure}

The seed embedding vocabulary captures intrinsic syntactic and semantic relations at the entity level of the LLVM IR (Sec.~\ref{subsection:seedEmbeddings-eval}). Hence, we believe that the primary strength of our encodings comes from the \textit{seed embedding vocabulary}. This directly leads to the \textit{Symbolic} encodings that achieve better performance than the other earlier methods on average.
When the \textit{Symbolic} encodings are augmented with flow information of the program, it results in \textit{Flow-Aware} encodings. These encodings result in much more informative representation and hence lead to better accuracy and speedup than all other methods. This is evident from the results shown in Sec.~\ref{sec:experimentation}.

However, this improvement in accuracy comes with a minimal overhead. Generating the \textit{Flow-Aware} encodings take more time than \textit{Symbolic} encoding, as it accounts for the time taken to generate and propagate the program flow information and to resolve circular dependencies in this process.
We did an analysis of time taken by both of these encodings on a sample set of straightforward programs involving the family of sorting, searching, dynamic and greedy programs obtained from an online collection of programs~\cite{geeksforgeeks}. 

The comparison of time taken to generate the \textit{Symbolic} and \textit{Flow-Aware} encodings from the \textit{Seed Embedding Vocabulary} on the sample set is shown in Fig.~\ref{fig:timetaken}. It can be observed that, on an average, \textit{Flow-Aware} encodings take $1.86$ times more than that of \textit{Symbolic} encodings. 
Their memory representations are of the same size, as both of them result in a floating-point vector in the same n-dimensions (300 dimensions for our setting).

\subsection{Exposure to OOV words}

For learning the representations of programs, the training phase of any method often involves learning a vocabulary that contains the embeddings corresponding to the input.
When an unseen combination of underlying input constructs is encountered during inference, it would not be a part of the vocabulary and leads to \textit{Out Of Vocabulary} (\textit{OOV}) data points. In such cases, most of the models treat all the \textit{OOV} words in a similar manner by assigning a common representation ambiguously, which may result in performance degradation.

Hence, to avoid \textit{OOV} points, it is important to expose various and large (all possible) combinations of the underlying entities during the training phase. For example, for generating the embeddings at a statement-level of IR, all combinations of opcodes, types and arguments that can potentially form a statement should be exposed during training.
Similarly, for generating token-based embeddings, all possible tokens that can possibly be encountered must have been exposed at the training time.
But, both of these approaches would lead to a huge training space of $\mathcal{O}(\lvert opcodes \rvert \times \lvert types \rvert \times \lvert arguments \rvert)$ and $\mathcal{O}(\lvert tokens \rvert)$ (where $ \lvert tokens \rvert$ can potentially be unbounded, with tokens being used more in the sense of a lexeme~\cite{Aho:2006:CPT:1177220}) respectively.
As it can be seen, covering such a huge intractable space is infeasible, and hence undesirable.

\begin{table}
  \caption{Comparison Matrix: DeepTune vs. NCC vs. \irtovec}
  \label{tab:comparisonMatrix}   
  \small
  \vspace*{-\baselineskip}
  \begin{tabularx}{\textwidth}{lllll}
    \toprule
     \textbf{Comparison metric} & \textbf{DeepTune~\cite{cummins2017end2end}} & \textbf{NCC~\cite{ncc}} & \textbf{\irtovec} \\
    \hline
    \textbf{Primary embedding} & Token and Character level & Instruction level & Entity level \\
    \textbf{Files examined} & Handpicked vocabulary & 24,030 & 13,029 \\
    \textbf{Vocabulary size} & 128 symbols & 8,565 statement embeddings & 64 entity embeddings \\
    \textbf{Entities examined} & Application specific & $\approx$640M XFG Statement Pairs & $\approx$134M Triplets \\
 \textbf{Vocabulary training} & \multirow{2}{*}{Task dependent} & \multirow{2}{*}{200 hrs on a P100} & \multirow{2}{*}{20 mins on a P100} \\
 \textbf{time} &  &  &\\
\bottomrule
\end{tabularx}
\vspace*{-\baselineskip}
\end{table}

Consequently, the methods that learn statement level representations like NCC~\cite{ncc}  face \textit{OOV} issues. However, DeepTune~\cite{cummins2017end2end} uses a Token and Character based hybrid approach to overcome this issue. DeepTune\textquotesingle s method involves usage of the embeddings corresponding to the token if it is present in vocabulary. Else, they break the token as a series of characters and use the corresponding embeddings of the character.

On the other hand, \irtovec forms embeddings at the entity level of the IR, and hence it is sufficient to expose a training space of only $\mathcal{O}(\lvert opcodes\rvert + \lvert types \rvert + \lvert arguments \rvert)$ to avoid \textit{OOV} points. 
With this insight, it can be seen that \irtovec can avoid the \textit{OOV} issue even on exposure to a smaller number of programs at training time, when compared to the other approaches. Consequently, this results in a smaller vocabulary, and hence achieving better performance than the other methods. 
A comparison of \irtovec with DeepTune and NCC with respect to training and vocabulary is shown in Tab.~\ref{tab:comparisonMatrix}.

\begin{figure}
    \centering
    \includegraphics[scale=0.34, width=\textwidth]{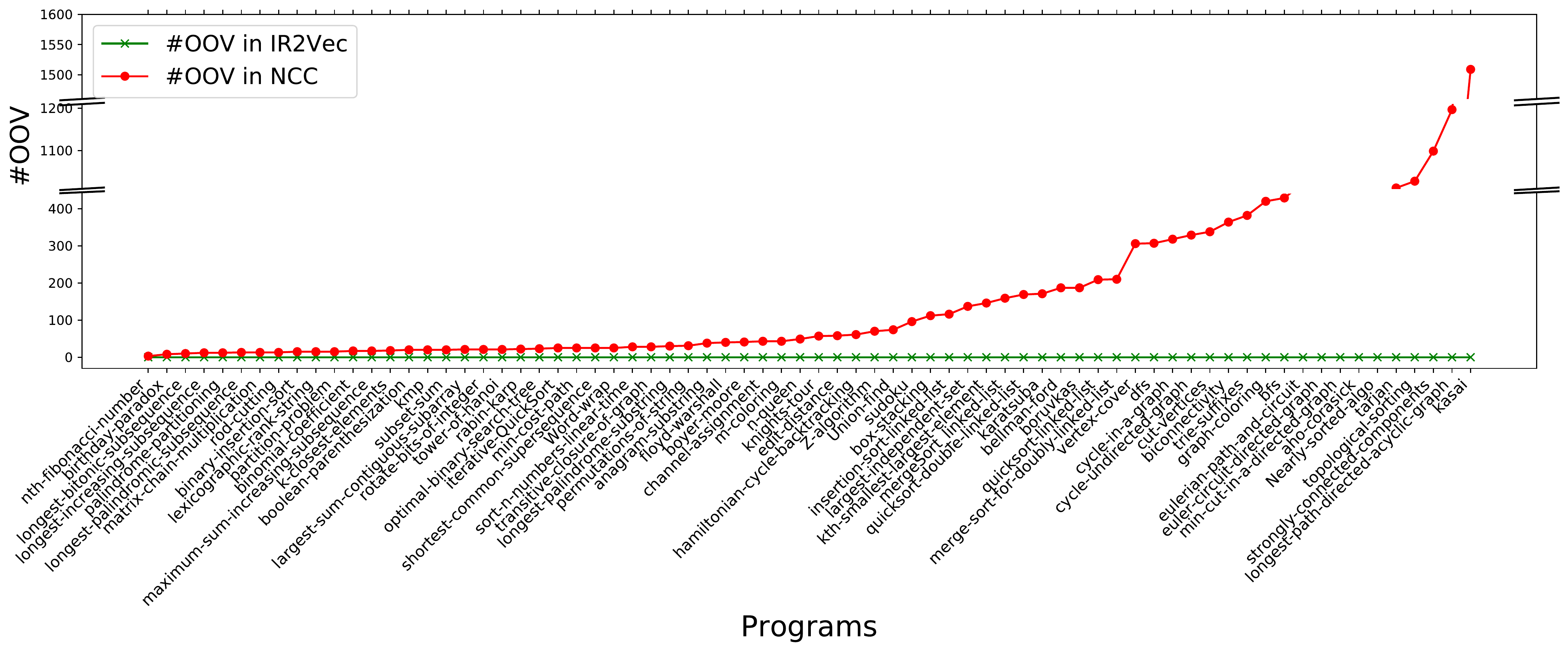} 
    \vspace{-\baselineskip}
    \caption{Comparison of the number of \textit{OOV} entities encountered by NCC and \irtovec}
    \label{fig:scalability}
     \vspace*{-0.5cm}
\end{figure}

A comparison of the number of \textit{OOV} entities encountered by NCC and \irtovec on the same set of programs used in Sec.~\ref{subsection:symbolicVsFlowaware} is shown in Fig.~\ref{fig:scalability}.
It can be seen that our method does not encounter any \textit{OOVs} even when exposed to lesser training data, thereby achieving good scalability. 

\section{Conclusions and Future Work} 
\label{sec:conclusions}
We proposed \irtovec, a novel LLVM-IR based framework that can capture the implicit characteristics of the input programs in a task-independent manner.
The seed embeddings were formed by modelling IR as relations, and the encoding was obtained by using a translational model. This encoding was combined with liveness, use-def, and reaching definition information, to form vectors at various levels of the program abstraction like instruction, function and module. Overall, this results in two encodings, which we term as \textit{Symbolic} and \textit{Flow-Aware}.

When compared to earlier approaches, our approach of representing programs is \textit{non data-hungry}, takes \textit{less training} time of up to $8640 \times$, while maintaining a \textit{small vocabulary} of \textit{only 64 entities}.
As we use entity level seed embeddings, we \textit{do not} encounter any \textit{OOV} issues. 
We demonstrate the effectiveness of the obtained encodings on two different tasks and obtain \textit{superior} performance results while achieving \textit{high} scalability when compared with various similar approaches. 

We envision that our framework can be applied to other applications beyond the scope of this work. \irtovec can be extended to classify whether a program is malicious or not by looking for suspicious and obfuscated patterns. It can also be applied for detecting codes with vulnerabilities, and to identify the patterns of code and replace them with its optimized equivalent library calls.
It can even be extended to aide in key optimizations like the prediction of vectorization, interleaving, and unrolling factors.
We also plan to extend the device mapping and thread coarsening experiments with more datasets and on newer platforms.

The source code and other relevant material are available in \url{http://www.compilers.cse.iith.ac.in/research/ir2vec}.
\begin{acks}
We are grateful to Suresh Purini, Dibyendu Das, Govindarajan Ramaswamy and Albert Cohen, for their valuable feedback on our work at various stages. 
We also thank Swapnil Dewalkar, Akash Banerjee and Rahul Utkoor for the thoughtful discussions in the early stages of this work. 
We would like to thank the anonymous reviewers of ACM TACO for their insightful and detailed comments which helped in improving the paper.

This research is funded by the Department of Electronics \& Information Technology and the Ministry of Communications \& Information Technology, Government of India. 
This work is partially supported by a Visvesvaraya PhD Scheme under the MEITY, GoI (PhD-MLA/04(02)/2015-16), an NSM research grant (MeitY/R\&D/HPC/2(1)/2014), a Visvesvaraya Young Faculty Research Fellowship from MeitY (MEITY-PHD-1149), and a faculty research grant from AMD.

\end{acks}

\bibliographystyle{alpha}
\bibliography{references}

\end{document}